\pdfoutput=1
\documentclass[11pt]{article}
\usepackage[pdftex]{graphicx,color} 
\usepackage{jheppub}
\usepackage{amsmath}
\usepackage{amssymb}
\usepackage{comment}
\usepackage{multirow}
\usepackage{mathtools}
\usepackage{dsdshorthand}
\usepackage{shuffle}
\newcommand{\bea}{\begin{equation}\begin{aligned}}
\newcommand{\eea}[1]{\label{#1}\end{aligned}\end{equation}}
\newcommand{\beq}{\begin{equation}}
\newcommand{\eeq}{\end{equation}}

\newcommand   \dd  {d}
\newcommand   \ii  {i}

\newcommand{\es}[2] {\begin{equation} \label{#1} \begin{split} #2 \end{split} \end{equation}}

\usepackage{tikz}
\usetikzlibrary{arrows,calc,shapes,decorations.pathmorphing,decorations.markings}
\tikzset{
>=stealth',
help lines/.style={dashed, thick},
axis/.style={<->},
important line/.style={thick},
connection/.style={thick, dotted},
  cross/.style={
    cross out,
    draw=black, 
    minimum size=5pt, 
    inner sep=0pt,
    outer sep=0pt
  },
->-/.style={decoration={
  markings,
  mark=at position #1 with {\arrow{>}}},postaction={decorate}}
}

\title{The type IIA Virasoro-Shapiro amplitude in $\text{AdS}_4\times\text{CP}^3$ from ABJM theory}
\author[a]{Shai M. Chester,}
\author[b]{Tobias Hansen}
\author[a]{and De-liang Zhong}

\affiliation[a]{Blackett Laboratory, Imperial College, Prince Consort Road, London, SW7 2AZ, UK}
\affiliation[b]{Department of Mathematical Sciences, Durham University,
Stockton Road, Durham, DH1 3LE, UK}
\emailAdd{s.chester@imperial.ac.uk}
\emailAdd{tobias.p.hansen@durham.ac.uk}
\emailAdd{d.zhong23@imperial.ac.uk}

\abstract{
We consider tree level scattering of gravitons in type IIA string theory on $AdS_4\times \mathbb{CP}^3$ to all orders in $\alpha'$, which is dual to the stress tensor correlator in $U(N)_k\times U(N)_{-k}$ ABJM theory in the planar large $N$ limit and to all orders in large $\lambda\sim N/k$. The small curvature expansion of this correlator, defined via a Borel transform, is given by the flat space Virasoro-Shapiro amplitude plus AdS curvature corrections. We fix curvature corrections by demanding that their resonances are consistent with the superconformal block expansion of the correlator and with a worldsheet ansatz in terms of single-valued multiple polylogarithms. The first correction is fully fixed in this way, and matches independent results from integrability, as well as the $R^4$ correction at finite AdS curvature that was previously fixed using supersymmetric localization. We are also able to fix the second curvature correction by using a few additional assumptions, and find that it also satisfies various non-trivial consistency checks. We use our results to fix the tree level $D^4R^4$ correction at finite AdS curvature, and to give many predictions for future integrability studies.
}

\begin{document}
\maketitle

\section{Introduction}

One of the outstanding questions in string theory is how to compute string scattering in the presence of Ramond-Ramond (RR) flux. This includes string theory in AdS, which is famously dual to conformal field theories (CFTs) in one less dimension \cite{Maldacena:1997re}. The textbook RNS prescription for the string worldsheet does not work in this case, while the pure spinor approach \cite{Berkovits:2000fe} has not been developed to the point where it can compute string scattering in practice. 

Instead, progress has recently been made in \cite{Alday:2022uxp,Alday:2022xwz,Alday:2023jdk,Alday:2023mvu} for tree level string scattering in type IIB string theory in $AdS_5\times S^5$ by combining two ingredients. Recall that the stress tensor multiplet correlator in $SU(N)$ $\mathcal{N}=4$ super-Yang-Mills (SYM) is dual to graviton scattering in $AdS_5\times S^5$, where the flat space limit is given by a Borel transform defined in \cite{Penedones:2010ue}. The first assumption is that AdS curvature corrections can be obtained by applying this same Borel transform to the CFT correlator, after suitably rescaling the Mellin space expression for this correlator by $\lambda\equiv g_\text{YM}^2N$ and expanding at large $\lambda$ in the planar large $N$ limit. In particular, one can apply this AdS curvature expansion to the superblock expansion of the correlator, which includes operators with large scaling dimensions that are dual to massive string states in the bulk. The second assumption is that the AdS curvature corrections to the correlator are given by a worldsheet ansatz in terms of single-valued multiple polylogarithms (SVMPLs).\footnote{That curvature corrections to all orders can be expressed in terms of SVMPLs was also recently confirmed \cite{Alday:2024rjs} in a case where the worldsheet theory is under full control: string theory on $AdS_3 \times \cal{M}$ with pure NSNS fluxes.} By combining these two assumptions, as well as crossing symmetry, the first AdS curvature correction was completely fixed, and matched previous results from integrability \cite{Gromov:2011de,Basso:2011rs,Gromov:2011bz} for the scaling dimensions of massive string operators, as well as the low energy expansion at finite AdS curvature as fixed from supersymmetric localization \cite{Binder:2019jwn,Chester:2020dja}. By further inputting results from integrability, the second AdS curvature correction was also fixed.\footnote{A similar strategy was also applied to open string scattering in type IIB string theory in the presence of D7 branes \cite{Alday:2024yax,Alday:2024ksp}, and type IIB string theory on $AdS_3\times S^3\times M_4$ for $M_4=K3, T^4$ \cite{Chester:2024wnb}.}

In this paper, we extend this story to type IIA string scattering in $AdS_4\times \mathbb{CP}^3$, which is dual to the 3d ABJM CFT with gauge group $U(N)_k\times U(N)_{-k}$ \cite{Aharony:2008ug}. In particular, the string coupling $g_s$, the string length $\ell_s = \sqrt{\alpha'}$, and the AdS radius $R$ are related to CFT parameters as
 \cite{Aharony:2008ug,Chester:2018aca,Aharony:2009fc,Bergman:2009zh}
\es{class}{
\nu\equiv \frac{R^4}{\ell_s^4} = 2\pi^2({N/k}-1/24) +\dots\,,\qquad g_s^2 =\frac{512(N/k)^2}{3c_T}+\dots\,,
}
where $c_T\sim N^{3/2}$. The type IIA limit is then given by large $c_T$ with $\nu$ fixed, such that the string coupling $g_s$ is small in this expansion. We consider the stress tensor multiplet correlator dual to graviton scattering in the planar limit to all orders in $1/\nu$, and in a small AdS curvature expansion we define as follows. Starting from the Mellin space expression $M_i(s,t)$ for this correlator, we define the AdS amplitude using the flat space limit formula \cite{Penedones:2010ue,Binder:2019mpb}\footnote{We include the shifts by $4/3$ since the Mellin variables satisfy $s+t+u=4$, but the Mandelstam variables should satisfy $S+T+U=0$.}
\bea
A_i(S,T) &= 
\frac{\sqrt{\pi}}{4 \sqrt{\nu}}
\int_{\kappa-i\infty}^{\kappa+ i \infty} \frac{d\alpha}{2 \pi i} \, e^\alpha \alpha^{-\frac12} M_i\left(\frac{2 \sqrt{\nu}}{\alpha} S + \frac43,  \frac{2 \sqrt{\nu}}{\alpha} T + \frac43 \right) \,,
\eea{FlatLimit}
where $S,T$ are the flat space Mandelstam variables
\beq
S = - \frac{\alpha'}{4} (p_1 + p_2)^2 \,, \qquad
T = - \frac{\alpha'}{4} (p_1 + p_4)^2 \,, \qquad
U = - \frac{\alpha'}{4} (p_1 + p_3)^2\,,
\eeq
which satisfy $S+T+U=0$. We consider the AdS amplitude in a small  curvature expansion in $1/\sqrt{\nu} = \alpha'/R^2$
\beq
A(S,T) = A^{(0)}(S,T) + \frac{1}{\sqrt{\nu}} A^{(1)}(S,T) +  \frac{1}{\nu} A^{(2)}(S,T) + \ldots\,,
\label{A_expansion}
\eeq
where we are still sensitive to finite $\alpha'$ effects via the Mandelstam variables, and the first term is the flat space AdS Virasoro-Shapiro amplitude
\beq
A^{(0)}(S,T)=\frac{\Gamma(1-S)\Gamma(1-T)\Gamma(1-U)}{\Gamma(1+S)\Gamma(1+T)\Gamma(1+U)}
\left( \frac{T U}{S} , \frac{S T}{U}, \frac{S U}{T}, \frac{S}{2}, \frac{U}{2}, \frac{T}{2}\right)\,.
\label{flat_VS}
\eeq
We then apply the Borel transform \eqref{FlatLimit} to the superblock expansion of the correlator \cite{Binder:2020ckj}, and compare it to an ansatz for $A^{(k)}(S,T)$ in terms of SVMPLs of weights up to $3k$. We find that $A^{(1)}(S,T)$ is completely fixed and can be written with an integrand of uniform weight 3. Our result can be checked against integrability results for massive string operators on the leading Regge trajectory \cite{Gromov:2014eha}, as well as the $R^4$ higher derivative correction at finite AdS curvature as fixed by analytic bootstrap combined with localization \cite{Binder:2019mpb}. We also find that $A^{(1)}(S,T)$ in the high energy limit matches the classical solution for the string scattering problem in AdS \cite{Alday:2023pzu}, and in particular the high energy limit of the $AdS_5\times S^5$ \cite{Alday:2023pzu} and $AdS_3\times S^3\times M_4$ \cite{Chester:2024wnb} cases.

To completely fix the second curvature correction $A^{(2)}(S,T)$ we have to make a few more assumptions. We assume that the worldsheet integrand has uniform weight 6 and that the leading Regge trajectories are non-degenerate. The superprimaries of exchanged long multiplets can be distinguished by their conformal dimension and spin as well as their parity $\pm$ under the $\mathbb{Z}_2$ symmetry $\cZ$ described in \cite{Binder:2020ckj}, which is a symmetry of ABJM theory,\footnote{We do not distinguish operators by their spacetime parity $\cP$, even though $U(N)_k\times U(N)_{-k}$ ABJM theory is parity invariant, because we are not sensitive to this parity in the planar limit.} so that we denote the leading Regge trajectories as $(\text{odd spin})_+$, $(\text{even spin})_+$ and $(\text{even spin})_-$. We then impose the dimension of the first operator on the $(\text{odd spin})_+$ trajectory from integrability, and the $R^4$ term known from localization. The consistency checks for the second curvature correction include a match with integrability data where available, i.e.\ for all operators on the $(\text{odd spin})_+$ trajectory \cite{Gromov:2014eha} and the first operator on the $(\text{even spin})_+$ trajectory \cite{Bombardelli:2018bqz}, a match with localization for the $D^4R^4$ correction and a match with the classical high energy limit which confirms the expected exponentiation property.
Our result implies the following conformal dimensions for spin $\ell$ operators with $\cZ$ parity $\pm$ on the leading Regge trajectories\footnote{In order to compare to \cite{Gromov:2014eha,Bombardelli:2018bqz} note that these references considered operators of dimension $\Delta + 1$ and spin $S = \ell+1$ which are super-descendants of the superconformal primaries of spin $\ell$ considered here.
Explicitly, $\Delta^{\text{odd } \ell}_+$ and $\Delta^{\text{even } \ell}_+$ match the scaling dimensions of operators within the $\mathfrak{sl}(2)$ subsector \cite{Beisert:2010jr,Klose:2010ki}. These operators take the form $\text{tr}\,D_{+}^S(Y^1 Y_4^{\dagger})^{L=1}$, where $S$ is even or odd, respectively. Currently, no integrability data is available for $\Delta^{\text{even } \ell}_-$.
}
\es{deltas}{
\Delta^{\text{odd } \ell}_+ ={}& - \frac32 +  \sqrt{2(\ell+1)} \nu^\frac14 \bigg[
1 + \frac{1}{\sqrt{\nu}} \bigg(  \frac{6 \ell^2+8 \ell+11}{16 (\ell+1)} \bigg)   \\
&-\frac{1}{512\nu} \bigg(\frac{84 \ell^4+208 \ell^3-12 \ell^2-384 \ell-167}{(\ell+1)^2}+288 (\ell+1) \zeta (3)   \bigg) + O(\nu^{-\frac32})
\bigg]\,, \\
\Delta^{\text{even } \ell}_+={}& - \frac32 +  \sqrt{2(\ell+2)} \nu^\frac14 \bigg[
1 + \frac{1}{\sqrt{\nu}} \bigg( \frac{6 \ell^2+16 \ell+9}{16( \ell+2)} \bigg)   \\
&-\frac{1}{512\nu} \bigg( \frac{84 \ell^4+480 \ell^3+940 \ell^2+672 \ell+81}{(\ell+2)^2}+288 (\ell+2) \zeta (3) \bigg) + O(\nu^{-\frac32})
\bigg]\,, \\
\Delta^{\text{even } \ell}_- ={}& - \frac32 +  \sqrt{2(\ell+2)} \nu^\frac14 \bigg[
1 + \frac{1}{\sqrt{\nu}} \bigg( \frac{6 \ell^2+16 \ell+21}{16( \ell+2)} \bigg)   \\
&-\frac{1}{512\nu} \bigg( \frac{84 \ell^4+480 \ell^3+796 \ell^2-48 \ell-615}{(\ell+2)^2}+288 (\ell+2) \zeta (3) \bigg) + O(\nu^{-\frac32})
\bigg]\,. \\
}

The rest of the paper is organised as follows.
Section \ref{sec:correlator} introduces the correlator of the stress-tensor multiplet in ABJM theory. In section \ref{sec:3} we relate the superconformal OPE for long supermultiplets to massive poles of the AdS amplitude and discuss the spectrum of massive string operators. Section \ref{sec:worldsheet} describes our ansatz for curvature corrections to the AdS amplitude as a worldsheet integral. In section \ref{sec:A1} and \ref{sec:A2} we fix the first two curvature corrections and present resulting data and checks, namely OPE data, Wilson coefficients and the high energy limit. We conclude in section \ref{sec:conclusions}.
Four appendices contain further technical details.

\section{Stress-tensor correlator}
\label{sec:correlator}

We study the four-point function of the superconformal primary $S$ of the stress-tensor multiplet, which is a $\Delta=1$ scalar in the adjoint of the $SO(6)$ R-symmetry. We can expand this correlator as
\begin{align}
\langle S(\vec x_1,X_1) \cdots S(\vec x_4,X_4)\rangle ={}& \frac 1{x_{12}^2 x_{34}^2}\Bigg[\mathcal{S}_1(U,V)A_{12}A_{34}+\mathcal{S}_2(U,V)A_{13}A_{24}+\mathcal{S}_3(U,V)A_{14}A_{23}
\nonumber \\
& + \mathcal{S}_4(U,V)B_{1423} + \mathcal{S}_5(U,V)B_{1234}+\mathcal{S}_6(U,V)B_{1342}\Bigg] \,,
\label{SSSScor}
\end{align}
where we define the R-symmetry structures
\begin{equation}\label{ABRDef}
A_{ij} = \Tr(X_iX_j) \,, \qquad B_{ijkl} = \Tr(X_i X_j X_k X_l) + \Tr(X_l X_k X_j X_i)  \,, 
\end{equation}
and where $\mathcal{S}_i$ are functions of the conformal cross-ratios
\begin{equation} U \equiv \frac{x_{12}^2x_{34}^2}{x_{13}^2x_{24}^2} \,, \qquad  V \equiv \frac{x_{14}^2x_{23}^2}{x_{13}^2x_{24}^2} \,.
\end{equation}
We will be interested in the connected part of this correlator 
\begin{equation}\mathcal{S}_i^{\text{conn}}(U,V)\equiv \mathcal{S}_i(U,V)-\mathcal{S}_i^{\text{disc}}(U,V) \,,  \qquad
 \mathcal{S}_i^{\text{disc}} = \begin{pmatrix} 1, & U, & \frac UV, &0,&0,&0\end{pmatrix} \,,
 \end{equation}
for which we define the Mellin transform (where $u=4-s-t$)\footnote{Note that we switch $t\leftrightarrow u$ relative to \cite{Binder:2019mpb}.}
\begin{equation}\begin{split}\label{mellin}
\mathcal{S}_i^{\text{conn}}(U,V)=&\int_{-i\infty}^{i\infty}\frac{ds\, dt}{(4\pi i)^2}\ U^{\frac s2}V^{\frac t2-1} \Gamma^2\left[1-\frac s2\right]\Gamma^{2}\left[1-\frac t2\right]\Gamma^{2}\left[1-\frac u2\right] M_i(s,t)\,.
\end{split}\end{equation}
The Mellin amplitudes satisfy the following crossing symmetry constraints, which imply that all components are fixed in terms of $M_2(s,t)$ and $M_5(s,t)$
\bea
M_1(s,t) &= M_2(u,t)\,, \qquad M_2(s,t) = M_2(t,s)\,, \qquad M_3(s,t) = M_2(s,u)  \,, \\
M_4(s,t) &= M_5(u,t)\,, \qquad M_5(s,t) = M_5(t,s)\,, \qquad M_6(s,t) = M_5(s,u) \,.
\eea{M_crossing}

\subsection{Mellin amplitudes for the low energy effective theory}

The Mellin amplitude can also be expanded at large $N\sim c_T^{2/3}$ and large $\nu\sim\lambda$ as\footnote{We normalised our Mellin amplitude with an extra factor of $32/\pi^2$ relative to \cite{Binder:2019mpb} to make various formulas simpler. This overall normalisation is not important as we only consider tree level.}
\beq
M = \frac{1}{c_T}M_\text{tree}+O(c_T^{-2})\,,
\eeq
with
\bea
M_\text{tree} ={}&M^\text{SG}+\frac{3\zeta(3)}{32\nu^{\frac32}}[35M^{4,1}-72M^2]\,,\\
&+\frac{1}{\nu^{\frac52}}\Big[\sum_{j=1}^3 b_{6,j} M^{6,j}+\sum_{j=1}^2 b_{5,j} M^{5,j}+\sum_{j=1}^2 b_{4,j} M^{4,j}+b_2M^2\Big]+O(\nu^{-3})\,.
\eea{Mexp}
where the $b$'s are numbers that were not yet fixed. The supergravity term was fixed by analytic bootstrap in \cite{Zhou:2017zaw,Binder:2019mpb} to take the form
\bea
M_2^\text{SG} (s,t)&=-\frac{(s-2)(t-2)}{u(u+2)}\left(\frac{4\Gamma\left(\frac{1-u}2\right)}{\sqrt{\pi}\Gamma\left(1-\frac u2\right)}-(4+u)\right)
\,,\\
M_5^\text{SG} (s,t)&= -\frac{u-2}{2s t}\left(\frac{2t\Gamma\left(\frac{1-s}2\right)}{\sqrt\pi \Gamma\left(1-\frac s2\right)}+\frac{2s\Gamma\left(\frac{1-t}2\right)}{\sqrt\pi \Gamma\left(1-\frac t2\right)}+2u-st-8\right)  \,,
\eea{M_SG}
where the other coefficients can be obtained by crossing using \eqref{M_crossing}. 
The ratio of gamma functions $\frac{\Gamma\left(\frac{1-s}2\right)}{\Gamma\left(1-\frac s2\right)}$ has a large $s$ expansion in terms of odd positive powers of $1/\sqrt{s}$, for all of which the Borel transform \eqref{FlatLimit} vanishes exactly. As a result, the curvature corrections to the supergravity AdS amplitude are all rational functions of the Mandelstams
\bea
A_2^\text{SG} (S,T)&= \frac{S T}{U} + \frac{1}{\sqrt{\nu}} \frac{S^2 + 3 S T + T^2}{6U^2} - \frac{1}{\nu} \frac{2S^2-13 S T + 2 T^2}{36 U^3} + O(\nu^{-\frac{3}{2}})\,,\\
A_5^\text{SG} (S,T)&= \frac{U}{2} - \frac{1}{\sqrt{\nu}} \frac{3S^2 + 7 S T + 3 T^2}{12 S T} + \frac{1}{\nu} \frac{U(2S^2- S T + 2 T^2)}{24 S^2 T^2} + O(\nu^{-\frac{3}{2}})\,.
\eea{A_SG}
The other Mellin amplitudes in \eqref{Mexp} correspond to higher derivative corrections to supergravity, and they are polynomials in $s,t$. For instance, the first correction shown in \eqref{Mexp} corresponds to the $R^4$ correction to supergravity, whose coefficients were fixed using supersymmetric localization in \cite{Binder:2019mpb}, where the Mellin amplitudes take the explicit form
\es{R4Mel}{
{M}^{4,1}_2 & = \frac{1}{35} (u-2) (t-2) \left(35 ts+100u-112 \right)\,,\\
  {M}^{4,1}_5 & = \frac 1 {70}(u-2)(35 s t u-90(t^2+s^2)-280 ts -324u+1072)\,, \\     
   {M}^{2}_2 & = (s-2)(t-2), \qquad  {M}^{2}_5  = (u-{4}/{3})(u-2)\,.
}
The corresponding AdS amplitude is given by
\begin{align}
A_2^{R^4}&= \zeta(3) \left(2 S^2 T^2 + \frac{23S T U}{3\sqrt{\nu}}  +  \frac{80 S^2+221 S T+80 T^2}{18\nu} + \frac{53U}{18\nu^{\frac{3}{2}}} + \frac{13}{108 \nu^2} \right)\,,
\label{A_R4}\\
A_5^{R^4}&=  \zeta(3) \left(S T U^2
- U\frac{41 S^2+105 S T+41 T^2}{6 \sqrt{\nu }}
+\frac{133 S^2+286 S T+133 T^2}{9 \nu }
 - \frac{647U}{72\nu^{\frac{3}{2}}} + \frac{20}{27 \nu^2}
\right)
\,.
\nonumber
\end{align}
The next correction corresponds to the $D^4R^4$ correction, and is written in terms of eight unknown coefficients $b$. In appendix \ref{app:mellin}, we describe how to derive the explicit polynomials in $s,t$ that appear at that order, and give explicit expressions for them. We will later fix these coefficients using the curvature corrections we will next discuss, together with one localization constraint.

\section{OPE and AdS resonances}
\label{sec:3}

\subsection{Superconformal block expansion}
\label{sec:OPE}

\begin{table}
\begin{center}
{\renewcommand{\arraystretch}{1.4}
\begin{tabular}{ l | c |   l|l }
Superconformal block    & {normalisation} & $\mathcal{P}$ & $\mathcal{Z}$   \\  \hline
\multirow{2}{*}{$\text{Long}_{\Delta, 0}^{n}$}  & $n=1:\quad(a_{\Delta, 0, {\bf 1}}, a_{\Delta+1, 0, {\bf 20'}})  = (1, 0)$ &  $+$ &  $+$    \\ 
 &    $n=2:\quad (a_{\Delta, 0, {\bf 1}}, a_{\Delta+1, 0, {\bf 20'}})  = (0, (\Delta+2)/(\Delta-1))$   &  $-$ &  $+$  \\ 
 \hline
$\text{Long}^1_{\Delta, \ell}$, $\ell \geq 1$ odd    &   $a_{\Delta+1, \ell+1, {\bf 15}_s} = 1$  & $+$ & $+$ \\ 
\hline
\multirow{3}{*}{$\text{Long}_{\Delta, \ell}^{n}$, $\ell \geq 2$ even}    & $n=1:\quad(a_{\Delta, \ell, {\bf 1}}, a_{\Delta+1, \ell, {\bf 1}}, a_{\Delta+1, \ell, {\bf 15}_s})  = (1, 0, 0)$ &  $+$ &  $+$  \\ 
  & $n=2:\quad(a_{\Delta, \ell, {\bf 1}}, a_{\Delta+1, \ell, {\bf 1}}, a_{\Delta+1, \ell, {\bf 15}_s})  = (0, 1, 0)$&  $-$ &  $+$ \\ 
&   $n=3:\quad(a_{\Delta, \ell, {\bf 1}}, a_{\Delta+1, \ell, {\bf 1}}, a_{\Delta+1, \ell, {\bf 15}_s})  = (0, 0, 1)$&  $-$ &  $-$  \\ 
 \hline
\end{tabular}}
\caption{A summary of the long superconformal blocks and their normalisations in terms of a few OPE coefficients.  The values $a_{\Delta', \ell', {\bf r}}$ in this table correspond to $a^{n,\Delta,\ell}_{\Delta',\ell',\mathbf{r}}$ in \eqref{N=6_block}. We omitted the labels $n,\Delta,\ell$ for clarity. }
\label{SupermultipletTable}
\end{center}
\end{table}

The correlator has the following expansion into $\mathcal{N}=6$ 
superconformal blocks
\beq
\mathcal{S}_i(U,V) = \sum\limits_{n,\tau,\ell}
C^2_{n,\tau,\ell} \text{Long}^n_{\Delta,\ell,i} (U,V) + \text{ short multiplets}\,,
\label{superOPE}
\eeq
where $\tau \equiv \Delta - \ell$ is the twist and $n$ denotes the different possible long blocks that appear for a given spin $\ell$.
We will only be interested in the contribution of long blocks, as these are the only ones contributing to the massive poles of the AdS amplitude.
The $\mathcal{N}=6$ long superconformal blocks are given in terms of bosonic $3d$ blocks $g_{\Delta,\ell} (U,V)$ \cite{Binder:2020ckj}
\beq
\text{Long}^n_{\Delta,\ell,i} (U,V) = \sum_{\Delta',\ell',\mathbf{r}} a^{n,\Delta,\ell}_{\Delta',\ell',\mathbf{r}} (\mathbf{B}^{-1})^{\mathbf{r}}_{\ i} g_{\Delta',\ell'} (U,V)\,,
\label{N=6_block}
\eeq
where the matrix $\mathbf{B}$ implements a basis change from the irreducible representations $\mathbf{r}$ to the basis \eqref{SSSScor}. The relative coefficients $ a^{n,\Delta,\ell}_{\Delta',\ell',\mathbf{r}}$ between bosonic blocks within the superblock are completely fixed by supersymmetry, and given explicitly in \cite{Binder:2020ckj}.\footnote{For $\ell=0$, we normalise the $n=2$ block differently than  \cite{Binder:2020ckj}, see table \ref{SupermultipletTable}.} The index $n$ refers to the fact that there are different long blocks for even $\ell$ (unlike odd $\ell$ which has a unique block), as shown in Table \ref{SupermultipletTable}. These different $n$ correspond to different charges under the discrete symmetries $\mathcal{P}$ and $\mathcal{Z}$.

As discussed in Appendix~B.3 of~\cite{Binder:2019mpb}, the stress-tensor multiplet forms a representation not only of the superconformal group $OSp(6|4)$, but also of a larger group $(\mathbb Z_2\times\mathbb Z_2)\ltimes OSp(6|4)$ which includes both a parity transformation $\cP$ and discrete R-symmetry transformation $\cZ$. The parity transformation $\cP$ extends the spacetime symmetries from $Spin(3, 2) \cong Sp(4, \R)$ to $Pin(3,2)$, while $\cZ$ extends the R-symmetry group from $SO(6)$ to $O(6)$.  The $U(N)_k\times U(N+M)_{-k}$ ABJ(M) theories are expected to preserve $\mathcal{Z}$ symmetry in general, and $\cP$ for $M=0, k/2$. We can thus distinguish $\cZ$-even and $\cZ$-odd operators. However, since we are not sensitive to $M$ in the planar limit, we are not sensitive to $\cP$, so we expect that even spin, $\cZ$-even long multiplets will contribute to both $n=1$ and $n=2$ in this limit.

We would like to understand how these superconformal blocks contribute to the massive $S$-channel poles of the AdS amplitude. To this end, we have to perform the Mellin transform \eqref{mellin} and the Borel transform \eqref{FlatLimit} near the resonances of the AdS amplitude from massive string operators with twists $\tau = 2 \nu^\frac14 \tl \tau$, where $\tl \tau$ is finite at large $\nu$. In fact, the identification of Casimirs in the flat space limit
\beq
\Delta(\Delta-3) = R^2 m^2 = R^2\frac{4 \delta}{\alpha'}\,, \text{ as } R\to \infty\,,  
\eeq
fixes the leading term in the strong coupling expansion of the twists in terms of the string mass levels $\delta = 1,2,3,\ldots$
\beq
\tl\tau_{n,\delta,\ell}  = \sqrt{\delta}
+ O(\nu^{-\frac14})\,.
\label{leading_tau}
\eeq
We proceed at the level of bosonic blocks $g_{\Delta,\ell} (U,V)$. Their Mellin transform is given by
\beq
g_{\Delta,\ell} (U,V) = \int_{-i\infty}^{i\infty}\frac{ds\, dt}{(4\pi i)^2}\ U^{\frac s2}V^{\frac t2-1} \Gamma^2\left[1-\frac s2\right]\Gamma^{2}\left[1-\frac t2\right]\Gamma^{2}\left[1-\frac u2\right] \sum\limits_{m=0}^\infty \frac{\mathcal{Q}_{\tau,\ell, m}(t-2)}{s-\tau-2m}\,,
\eeq
where $\mathcal{Q}_{\tau,\ell, m}(t-2)$ is a Mack polynomial, defined in \eqref{curlyQ}, and $m$ labels descendants.
We show in appendix \ref{app:ads_poles} that applying the Borel transform to the Mellin amplitude
\beq
M_{\tau,\ell}(s,t) = \sum_{m=0}^\infty
\frac{\mathcal{Q}_{\tau,\ell,m}(t-2)}{s-\tau-2m}\,,
\label{Mtauell_app}
\eeq
leads to the following expansion for the AdS amplitude at $S \sim \tl \tau^2$
\beq
A_{\tau,\ell}(S,T)\Big|_{S\text{-poles}} = \frac{1}{16 \pi^\frac{5}{2} \nu^\frac{5}{8}} \sin\left( \frac{\pi\tau}2  \right)^2
\sum_{j=0}^\infty \frac{1}{\nu^{j/4}} R^{(j)}_{\tau,\ell}(S,T)\,,
\label{R_def}
\eeq
with
\beq
\begin{aligned}
R^{(0)}_{\tau,\ell}(S,T) ={}& -\frac{\sqrt{S} \sqrt{\tl{\tau}} P_\ell\left(1+\frac{2 T}{S}\right)}{\sqrt{2} \left(S-\tl{\tau}^2\right)}\,,\\
R^{(1)}_{\tau,\ell}(S,T) ={}&
-\frac{\sqrt{S} \left((4 \ell-1) S+(12 \ell-23) \tl{\tau}^2\right) P_\ell\left(1+\frac{2 T}{S}\right)}{16 \sqrt{2} \sqrt{\tl{\tau}}
   \left(S-\tl{\tau}^2\right)^2}\,,
\label{R_results}
\end{aligned}
\eeq
where $P_{\ell}\left( x\right)$ is a Legendre polynomial, as appropriate for partial waves in 4-dimensional flat space
and the higher corrections $R^{(2)}$ etc.\ can also depend on $\partial^n_x P_{\ell-n}\left( x\right)$ and are given explicitly in the attached \texttt{Mathematica} notebook.

In order to obtain the superconformal pole structure of $A_i(S,T)$, we have to replace each bosonic block in \eqref{N=6_block} with \eqref{R_def}, with the appropriate $\Delta'$ and $\ell'$. Since each 
superconformal block contains only bosonic blocks of the same twist modulo 2, the factor $\sin\left( \frac{\pi\tau}2  \right)^2$ in \eqref{R_def} can be factored out of the superconformal blocks.
Cancelling this factor we write the OPE coefficients as
\beq
C^2_{n=1,\tau,\ell} = \frac{16 \pi^\frac{5}{2} \left( 2 \sqrt{\delta} \nu^\frac{1}{4} \right)^{\frac52}}{\sin\left( \frac{\pi\tau}2  \right)^2} f_{n=1,\delta,\ell}\,, \qquad
C^2_{n=2,3,\tau,\ell} = \frac{16 \pi^\frac{5}{2} \left( 2 \sqrt{\delta} \nu^\frac{1}{4} \right)^{\frac52}}{\sin\left( \frac{\pi(\tau+1)}2  \right)^2} f_{n=2,3,\delta,\ell}\,.
\label{f_def}
\eeq

\subsection{OPE data at strong coupling}

Next we expand the OPE data at strong coupling. The leading twists are given by \eqref{leading_tau} in terms of the string mass level $\delta$, which we will use to label the OPE data alongside the spin $\ell$ and the superconformal block label $n$. We expand the twists as
\beq
\tau_{n,\delta,\ell}  = 
\tau^{(1)}_{n,\delta,\ell}
+ 2 \sqrt{\delta \sqrt{\nu}} \left(1
+ \frac{\tau^{(2)}_{n,\delta,\ell}}{\delta\sqrt{ \nu}} 
+ \frac{\tau^{(4)}_{n,\delta,\ell}}{(\delta \sqrt{\nu})^{2}}+ O(\nu^{-\frac32})
\right)\,,
\eeq
and the OPE coefficients \eqref{f_def}
\beq
f_{n,\delta,\ell}  =
f^{(0)}_{n,\delta,\ell} 
+  \frac{f^{(1)}_{n,\delta,\ell}}{(\delta \sqrt{\nu})^{\frac{1}{2}}} 
+  \frac{f^{(2)}_{n,\delta,\ell}}{\delta \sqrt{\nu}}
+  \frac{f^{(3)}_{n,\delta,\ell}}{(\delta \sqrt{\nu})^{\frac{3}{2}}}
+  \frac{f^{(4)}_{n,\delta,\ell}}{(\delta \sqrt{\nu})^{2}}
 + O(\nu^{-\frac54})\,.
\eeq
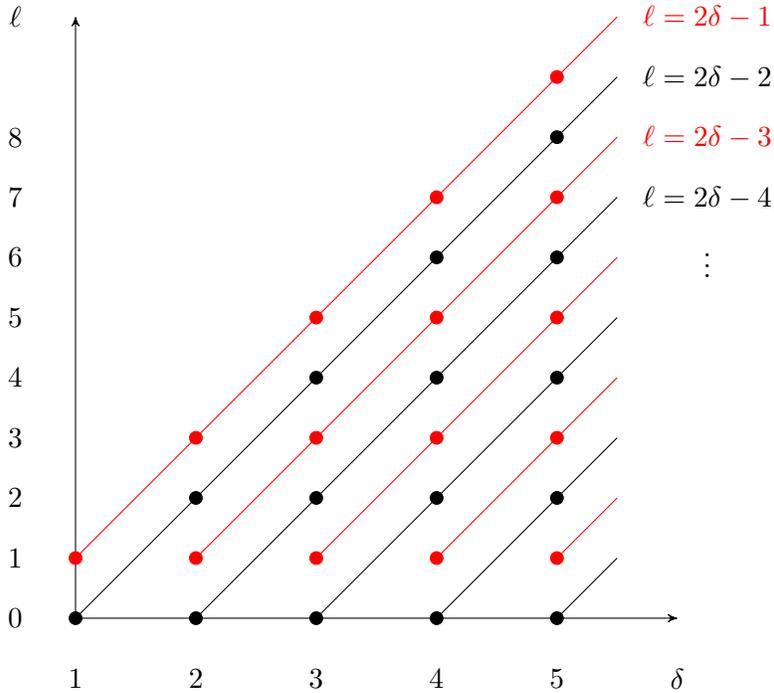
\begin{figure}[tb]
\centering
  \begin{tikzpicture}[scale=0.8]
    \coordinate (nw) at (0,10);
    \coordinate (sw) at (0,0);
    \coordinate (se) at (10,0);
    \draw[->] (sw) --  (nw) ;
    \draw[->] (sw) --  (se) ;
    \node at (0,-1) [] {$1$}; 
    \node at (2,-1) [] {$2$}; 
    \node at (4,-1) [] {$3$}; 
    \node at (6,-1) [] {$4$}; 
    \node at (8,-1) [] {$5$}; 
    \node at (10,-1) [] {$\delta$}; 
    \node at (-1,0) [] {$0$};
    \node at (-1,1) [] {$1$};
    \node at (-1,2) [] {$2$};
    \node at (-1,3) [] {$3$};
    \node at (-1,4) [] {$4$};
    \node at (-1,5) [] {$5$};
    \node at (-1,6) [] {$6$};
    \node at (-1,7) [] {$7$};
    \node at (-1,8) [] {$8$};
    \node at (-1,10) [] {$\ell$};
    \draw[-,black] (0,0) --  (9,9) ;
    \draw[-,black] (2,0) --  (9,7) ;
    \draw[-,black] (4,0) --  (9,5) ;
    \draw[-,black] (6,0) --  (9,3) ;
    \draw[-,black] (8,0) --  (9,1) ;
    \draw[-,red] (0,1) --  (9,10) ;
    \draw[-,red] (2,1) --  (9,8) ;
    \draw[-,red] (4,1) --  (9,6) ;
    \draw[-,red] (6,1) --  (9,4) ;
    \draw[-,red] (8,1) --  (9,2) ;
    \node at (10.5,10) [red] {$\ell = 2\delta-1$};
    \node at (10.5,8) [red] {$\ell = 2\delta-3$};
    \node at (10.5,9) [black] {$\ell = 2\delta-2$};
    \node at (10.5,7) [black] {$\ell = 2\delta-4$};
    \node at (10.5,6) [] {$\vdots$};
    \filldraw [black] (0,0) circle (3pt);
    \filldraw [black] (2,0) circle (3pt);
    \filldraw [black] (2,2) circle (3pt);
    \filldraw [black] (4,0) circle (3pt);
    \filldraw [black] (4,2) circle (3pt);
    \filldraw [black] (4,4) circle (3pt);
    \filldraw [black] (6,0) circle (3pt);
    \filldraw [black] (6,2) circle (3pt);
    \filldraw [black] (6,4) circle (3pt);
    \filldraw [black] (6,6) circle (3pt);
    \filldraw [black] (8,0) circle (3pt);
    \filldraw [black] (8,2) circle (3pt);
    \filldraw [black] (8,4) circle (3pt);
    \filldraw [black] (8,6) circle (3pt);
    \filldraw [black] (8,8) circle (3pt);
    \filldraw [red] (0,1) circle (3pt);
    \filldraw [red] (2,1) circle (3pt);
    \filldraw [red] (2,3) circle (3pt);
    \filldraw [red] (4,1) circle (3pt);
    \filldraw [red] (4,3) circle (3pt);
    \filldraw [red] (4,5) circle (3pt);
    \filldraw [red] (6,1) circle (3pt);
    \filldraw [red] (6,3) circle (3pt);
    \filldraw [red] (6,5) circle (3pt);
    \filldraw [red] (6,7) circle (3pt);
    \filldraw [red] (8,1) circle (3pt);
    \filldraw [red] (8,3) circle (3pt);
    \filldraw [red] (8,5) circle (3pt);
    \filldraw [red] (8,7) circle (3pt);
    \filldraw [red] (8,9) circle (3pt);
  \end{tikzpicture}
\caption{Chew-Frautschi plot of the stringy operators.} \label{fig:chew_frautschi}
\end{figure}

By matching the OPE poles of $A^{(0)}_i(S,T)$ to the flat space Virasoro-Shapiro amplitude \eqref{flat_VS}, we find the spectrum of exchanged supermultiplets displayed in Figure \ref{fig:chew_frautschi}, and we can also compute the leading OPE coefficients $\< f^{(0)} \>_{n,\delta,\ell}$.
While the amplitude fixes the OPE data for any $n,\delta,\ell$, the data organises itself into analytic Regge trajectories as indicated in the figure. As familiar from flat space, the leading Regge trajectory determines the Regge limit of the amplitude. This was also recently confirmed in the $AdS_5 \times S^5$ context in \cite{Alday:2024xpq}.
We will merely use the Regge trajectories as a useful way to present OPE data and show the data for the first few Regge trajectories.
We define the angle brackets as a sum over all operators that share the same $n$,  $\delta$ and $\ell$.
Defining
\bea
r^\text{odd}_j(\delta) &= 
\frac{(-1)^{\delta }  \delta ^{2 \delta -2 j+1} \Gamma \left(j-2 \delta -\frac{1}{2}\right)}{4^{\delta +4}\Gamma (\delta ) \Gamma \left(j-\delta -\frac{1}{2}\right)}\,, \\
r^\text{even}_j(\delta) &=
\frac{(-1)^{\delta }  \delta ^{2 \delta -2 j+1} (4 \delta -4 j+1) \Gamma \left(j-2 \delta +\frac{1}{2}\right)}{4^{\delta +4}\Gamma (\delta ) \Gamma \left(j-\delta
   +\frac{1}{2}\right)}\,,
\eea{r_def}
we have for the first few odd spin Regge trajectories
\bea
\< f^{(0)} _{1,\delta,2 \delta-1}\> &= 4 r^\text{odd}_1(\delta)\,,\\
\< f^{(0)} _{1,\delta,2 \delta-3}\> &=
\frac{2}{3} \left(2 \delta ^3+6 \delta ^2-17 \delta +6\right) r^\text{odd}_2(\delta)\,,\\
\< f^{(0)} _{1,\delta,2 \delta-5}\> &= \frac{2}{45} \left(5 \delta ^6+29 \delta ^5-58 \delta ^4-373 \delta ^3+722 \delta ^2-97 \delta -210\right)r^\text{odd}_3(\delta)\,,
\eea{f0_odd_results}
and for the even spin ones
\begin{align}
\< f^{(0)} _{1,\delta,2 \delta-2}\> + \frac32 \< f^{(0)} _{2,\delta,2 \delta-2}\>&=
\< f^{(0)} _{3,\delta,2 \delta-2}\> =
\frac{2}{\delta} r^\text{even}_1(\delta)\,,
\nonumber\\
\< f^{(0)} _{1,\delta,2 \delta-4}\> + \frac32 \< f^{(0)} _{2,\delta,2 \delta-4}\>&=
\< f^{(0)} _{3,\delta,2 \delta-4}\> =
\frac{2}{3} \left(\delta ^2+2 \delta -5\right) r^\text{even}_2(\delta)\,,
\label{f0_even_results}\\
\< f^{(0)} _{1,\delta,2 \delta-6}\> + \frac32 \< f^{(0)} _{2,\delta,2 \delta-6}\>&=
\< f^{(0)} _{3,\delta,2 \delta-6}\> =
\frac{1}{45} \left(5 \delta ^5+19 \delta ^4-29 \delta ^3-253 \delta ^2+249 \delta +189\right)r^\text{even}_3(\delta)\,.
\nonumber
\end{align}
Note that only a combination of the $n=1$ and $n=2$ coefficients for even spins are fixed, because the corresponding blocks are indistinguishable in the flat space limit. The individual coefficients will only be fixed when determining $A^{(1)}_i(S,T)$.

As indicated in \eqref{A_expansion}, we expect the AdS amplitude to have curvature corrections which multiply integer powers of $\alpha'/R^2 = 1 / \sqrt{\nu}$.
However, the OPE data is expanded in powers of $\nu^{-\frac14}$ and imposing \eqref{A_expansion} fixes part of the OPE data.
In particular, absence of an AdS amplitude at order $\nu^{-\frac14}$ (and its OPE poles) fixes the OPE data
\bea
\tau^{(1)}_{1,\delta,\ell} &= -\ell -\frac{3}{2}\,,\quad && \ell \text{ odd}\,,\\
2 \tau^{(1)}_{1,\delta,\ell} f^{(0)}_{1,\delta,\ell}
+ 3 \tau^{(1)}_{2,\delta,\ell} f^{(0)}_{2,\delta,\ell}&= 
\left(-\ell -\frac{3}{2} \right) 
\left(2  f^{(0)}_{1,\delta,\ell}
+ 3 f^{(0)}_{2,\delta,\ell}\right)\,,
\quad
\tau^{(1)}_{3,\delta,\ell} = -\ell -\frac{3}{2}\,,\quad && \ell \text{ even}\,,
\eea{tau1}
as well as
\bea
 f^{(1)}_{1,\delta,\ell} ={}& \frac{11}{16}  f^{(0)}_{1,\delta,\ell} \,, \quad && \ell \text{ odd}\,,\\
f^{(1)}_{1,\delta,\ell} + \frac32 f^{(1)}_{2,\delta,\ell} ={}& \frac{27}{16} f^{(0)}_{1,\delta,\ell} - \frac{15}{32} f^{(0)}_{2,\delta,\ell}\,, \quad
 f^{(1)}_{3,\delta,\ell} = \frac{11}{16}  f^{(0)}_{3,\delta,\ell} \,, \quad && \ell \text{ even}\,.
\eea{f1}
The remaining $\tau^{(1)}$'s and $f^{(1)}$'s are fixed at order $\nu^{-\frac34}$ such that
\bea
\tau^{(1)}_{n,\delta,\ell} &= -\ell -\frac{3}{2}\,,\\
 f^{(1)}_{1,\delta,\ell} &= \frac{27}{16}  f^{(0)}_{1,\delta,\ell} \,, \qquad
f^{(1)}_{2,\delta,\ell} = -\frac{5}{16}  f^{(0)}_{2,\delta,\ell}  \,, \quad \ell \text{ even}\,.
\eea{remaining_tau1_f1}
Similarly, absence of corrections of order  $\nu^{-\frac34}$ and $\nu^{-\frac54}$ fixes all $f^{(3)}$'s
\bea
 f^{(3)}_{1,\delta,\ell} ={}&
\frac{11}{16} (f^{(2)}_{1,\delta,\ell}-f^{(0)}_{1,\delta,\ell} \tau^{(2)}_{1,\delta,\ell})
+\frac{3 \left(256 \ell^2+768 \ell+557\right) f^{(0)}_{1,\delta,\ell}}{4096  }
\,, \quad &&
\ell \text{ odd}\,,\\
 f^{(3)}_{1,\delta,\ell} ={}&
\frac{27}{16} (f^{(2)}_{1,\delta,\ell}-f^{(0)}_{1,\delta,\ell} \tau^{(2)}_{1,\delta,\ell})
+\frac{\left(1280 \ell^2+1280\ell-3849\right) f^{(0)}_{1,\delta,\ell}}{4096  }
\,, \quad &&
\ell \text{ even}\,,\\
 f^{(3)}_{2,\delta,\ell} ={}&
-\frac{5}{16} (f^{(2)}_{2,\delta,\ell}-f^{(0)}_{2,\delta,\ell} \tau^{(2)}_{2,\delta,\ell})
+\frac{3 \left(256 \ell^2+256 \ell-675\right) f^{(0)}_{2,\delta,\ell}}{4096 }
\,,&&\\
 f^{(3)}_{3,\delta,\ell} ={}&
\frac{11}{16} (f^{(2)}_{3,\delta,\ell}-f^{(0)}_{3,\delta,\ell} \tau^{(2)}_{3,\delta,\ell})
+\frac{3 \left(256 \ell^2+256 \ell+45\right) f^{(0)}_{3,\delta,\ell}}{4096 }
\,.&&
\eea{f3}
We expect the relations determining $\tau^{(1)}_{n,\delta,\ell}$, $f^{(1)}_{n,\delta,\ell}$ and $f^{(3)}_{n,\delta,\ell}$ to hold for each individual operator, as indicated by the absence of angle brackets.

\subsection{Degeneracies from flat space}

Let us take this opportunity to discuss the degeneracy of operators with the labels $\delta, \ell, n$. We can address this question in the strict flat space limit, by relating the massive spectrum of type II string theory in 10 dimensions to the flat space limit of $AdS_4 \times \mathbb{C P}^3$. By continuity of the degeneracies the result should then also hold at finite $\nu$.
An analogous analysis was done in \cite{Alday:2023flc} for type IIB string theory in $AdS_5 \times S^5$. In that case, several of the degeneracies were confirmed to hold at finite $\nu$ by quantum spectral curve computations in \cite{Gromov:2023hzc}.
Here we will only discuss the leading odd and even Regge trajectories.

In the present case we have the additional complication of non-maximal supersymmetry. However, the flat space amplitude \eqref{flat_VS} respects $\mathcal{N}=8$ supersymmetry, as one can check by relating its massive poles to $\mathcal{N}=8$ superconformal blocks.
These are given in terms of the $\mathcal{N}=6$ superconformal blocks by \cite{Binder:2020ckj}
\begin{align}
\text{Long}^{\mathcal{N}=8}_{\Delta,\ell,i} &= 
\text{Long}_{\Delta,\ell,i}^{ 1}+\frac{4 (\ell-1)^2 (-\Delta +\ell+1) (\Delta +\ell)}{(2 \ell-1) (2 \ell+1) (\ell-\Delta ) (\Delta +\ell+1)}
\text{Long}_{\Delta+1,\ell-1,i}\nonumber\\
&   +\frac{4 \Delta  (-\Delta +\ell+1) (\Delta +\ell)}{3 (\Delta +2) (\ell-\Delta ) (\Delta +\ell+1)}
\text{Long}_{\Delta+1,\ell,i}^{2}
     +\frac{4 (-\Delta +\ell+1) (\Delta +\ell)}{(\ell-\Delta ) (\Delta +\ell+1)}
\text{Long}_{\Delta+1,\ell,i}^{3}\nonumber\\
&      +\frac{4 (\ell+1) (\ell+2) (\Delta +\ell) (\Delta +\ell+2)}{(2 \ell+1) (2 \ell+3) (\Delta +\ell+1) (\Delta +\ell+3)}
\text{Long}_{\Delta+1,\ell+1,i}\nonumber\\
&    +\frac{4 (\Delta +4)^2 (-\Delta +\ell+1) (\Delta +\ell)}{(2 \Delta +5) (2 \Delta +7) (\ell-\Delta ) (\Delta +\ell+1)}
\text{Long}_{\Delta+2,\ell,i}^{1}\,.
\label{N=8to6}
\end{align}
We find that
\beq
\Res_{S=\delta} A_i^{(0)} (S,T) = \sum\limits_{\ell=0,2,\ldots}^{2\delta-2} \left(C^{\mathcal{N}=8}_{\Delta,\ell}\right)^2 \text{Long}^{\mathcal{N}=8}_{\Delta,\ell,i}\Big|_\text{flat space limit}\,,
\eeq
where we take the flat space limit as described in section \ref{sec:OPE}.
This means that we can do the degeneracy analysis in two steps. We first relate the flat space spectrum to $\mathcal{N}=8$ superconformal multiplets. In a second step we decompose the $\mathcal{N}=8$ multiplets into $\mathcal{N}=6$ multiplets using \eqref{N=8to6}.

The main result from the first step is that the operators on the leading $\mathcal{N}=8$ Regge trajectory with superprimary spin $\ell=2\delta-2$ all have degeneracy 1.
We then see from \eqref{N=8to6} that $\mathcal{N}=8$ multiplets with $\ell=2\delta-2$ contribute to $\mathcal{N}=6$ multiplets with odd spin $\ell=2\delta-1$ or even spin $\ell=2\delta-2$ and $n=2,3$
only with a single term in \eqref{N=8to6}, so these $\mathcal{N}=6$ trajectories should have degeneracy 1 as well.

We conclude that there should be three Regge trajectories of unique operators: one for $\ell=2\delta-1$, one for $\ell=2\delta-2$ with $\cZ$-even operators and  one for $\ell=2\delta-2$ with $\cZ$-odd operators. Below we will find further evidence for this statement, as we will use it as an assumption to fix the last few coefficients in the ansatz for the second curvature correction. These coefficients will then be subject to non-trivial consistency checks against integrability and localization results.

\section{Worldsheet correlator}
\label{sec:worldsheet}

\subsection{Ansatz}

Next, we will try to determine the curvature corrections 
$A^{(1)}(S,T)$ and $A^{(2)}(S,T)$ in \eqref{A_expansion}, by making an ansatz for these functions in terms of a worldsheet integral.
This is very much analogous to the ansatz of \cite{Alday:2023jdk,Alday:2023mvu} for $AdS_5 \times S^5$. It was argued there that the worldsheet integral for the $k$th curvature correction should be of the form of the flat space Virasoro-Shapiro amplitude, with the additional insertion of single-valued multiple polylogarithms (SVMPLs) of maximal weight $3k$.

We will construct an ansatz for the structures 2 and 5. To this end we take $i=2,5$ henceforth, and construct functions satisfying
\beq
A_{i} (S,T) = A_{i} (T,S)\,, \quad i=2,5\,.
\label{crossing}
\eeq
The remaining structures are then fixed by \eqref{M_crossing}
\bea
A_1(S,T) &= A_2(U,T)\,, \qquad  A_3(S,T) = A_2(S,U)  \,, \\
A_4(S,T) &= A_5(U,T)\,, \qquad  A_6(S,T) = A_5(S,U) \,.
\eea{A_crossing}
The ansatz takes the form
\beq
A_{i} (S,T) =  B_{i,1} (S,T) + B_{i,2} (U,T) + B_{i,2} (S,U) + B_{i,3} (U,T) - B_{i,3} (S,U)\,, \quad i=2,5\,,
\eeq
where
\bea
B_{i,1} (S,T) &= B_{i,1} (T,S)\,,\\
B_{i,2} (S,T) &= B_{i,2} (T,S)\,,\\
B_{i,3} (S,T) &= -B_{i,3} (T,S)\,,
\eea{B_symmetry}
and
\beq
B_{i,j} (S,T) = \int d^2z |z|^{-2S-2} |1-z|^{-2T-2} G_{i,j}(S,T,z)\,.
\eeq
The integrand has an expansion analogous to \eqref{A_expansion}
\beq
G_{i,j}(S,T,z) = G^{(0)}_{i,j}(S,T,z)
+ \frac{1}{\sqrt{\nu}}G^{(1)}_{i,j}(S,T,z)
+ \frac{1}{\nu}G^{(2)}_{i,j}(S,T,z) + \ldots
\eeq
The main idea behind the ansatz is that it is manifestly crossing symmetric \eqref{crossing}, while at the same time $G^{(k)}_{i,j}(S,T,z)$ will depend on $z$ only via SVMPLs.
When we now write the AdS amplitude as a single integral
\beq
A^{(k)}_i (S,T) = \int d^2z |z|^{-2S-2} |1-z|^{-2T-2} G^{(k)}_{i,\text{tot}}(S,T,z)\,,
\label{Gtot_def}
\eeq
the necessary changes of variables introduce additional factors of $|z|^2$ or $|1-z|^2$
\bea
G^{(k)}_{i,\text{tot}}(S,T,z) ={}&
G^{(k)}_{i,1}(S,T,z)
+ |z|^2 \left( G^{(k)}_{i,2}(U,T,1/z) + G^{(k)}_{i,3}(U,T,1/z)\right)\\
&+ |1-z|^2 \left( G^{(k)}_{i,2}(S,U,\tfrac{z}{z-1}) - G^{(k)}_{i,3}(S,U,\tfrac{z}{z-1}) \right)\,, \quad i=2,5\,.
\eea{Gtot}
Using \eqref{A_crossing} we can also obtain the remaining integrands
\bea
G^{(k)}_{i-1,\text{tot}}(S,T,z) ={}&
 |z|^2 G^{(k)}_{i,1}(U,T,1/z)
+G^{(k)}_{i,2}(S,T,z) + G^{(k)}_{i,3}(S,T,z) \\
&+ |1-z|^2 \left( G^{(k)}_{i,2}(S,U,\tfrac{z}{z-1}) + G^{(k)}_{i,3}(S,U,\tfrac{z}{z-1}) \right)\,, \quad i=2,5\,,\\
G^{(k)}_{i+1,\text{tot}}(S,T,z) ={}&
|1-z|^2 G^{(k)}_{i,1}(S,U,\tfrac{z}{z-1})
+ |z|^2 \left( G^{(k)}_{i,2}(U,T,1/z) - G^{(k)}_{i,3}(U,T,1/z)\right)\\
&+G^{(k)}_{i,2}(S,T,z) - G^{(k)}_{i,3}(S,T,z)\,, \quad i=2,5 \,.
\eea{Gtot_1346}
For the integrand for the $k$th curvature correction we propose the following ansatz, satisfying \eqref{B_symmetry}
\bea
G^{(k)}_{i,j=1,2}(S,T,z) &= \sum_{w=0}^{3k} \sum_b \left(P^{+,k,i,j}_{w,b}(S,T) T^{+}_{w,b}(z) + P^{-,k,i,j}_{w,b}(S,T) T^{-}_{w,b}(z)
\right)\,,\\
G^{(k)}_{i,3}(S,T,z) &= \sum_{w=0}^{3k} \sum_b \left(P^{+,k,i,3}_{w,b}(S,T) T^{-}_{w,b}(z) + P^{-,k,i,3}_{w,b}(S,T) T^{+}_{w,b}(z)
\right)\,,
\eea{G_ansatz_symmetrized}
where $P^{\pm}_{w,b}(S,T)$ are (anti-)symmetric homogeneous polynomials of degree $2+w-k$. The degree of the polynomials is chosen such that for each $k$, the contribution to the supergravity amplitude \eqref{A_SG} will have transcendental weight 0.  $T^\pm_{w,b}(w)$ are a basis of (anti-)symmetric single-valued multiple polylogs (SVMPLs) of weight $w$, defined by
\bea
\mathcal{L}^\pm_{W}(z) ={}& 
\mathcal{L}_{W}(z) \pm \mathcal{L}_{W}(1-z)
+\mathcal{L}_{W}(\bar{z}) \pm \mathcal{L}_{W}(1-\bar{z})\,,
\eea{L_symmetrised}
and single-valued multiple zeta values (SVMZVs).
The SVMPLs $\mathcal{L}_{W}(z)$ were introduced in \cite{Brown:2004ugm}.
They depend on a word $W$ made from the letters $0$ and $1$, the locations of the singularities of the worldsheet integrand (along with $\infty$).
They are single-valued in $z$ and satisfy the relation
\beq
\partial_z \cL_{\sigma W} (z) = \frac{1}{z-\sigma} \cL_{W} (z)\,, \qquad \sigma \in \{0,1 \}\,.
\eeq
See \cite{Alday:2023jdk,Alday:2023mvu} for further details on SVMPLs in the same context and \cite{Duhr:2019tlz,HyperlogProcedures,Panzer:2014caa} for useful tools for manipulating them.
The explicit basis of SVMPLs we use is given in appendix \ref{app:SVMPLs} and we give a counting of the basis elements in table \ref{table:basis_dimensions}.
\begin{table}[h!]
\begin{center}
\begin{tabular}{ l | c c c c c c c } 
 $w$ & 0 & 1 & 2 & 3 & 4 & 5 & 6\\ 
 \hline
$P^{+}_{w,b}(S,T)$ & \multicolumn{7}{c}{$\lfloor \frac{4+w-k}{2} \rfloor$} \\
$P^{-}_{w,b}(S,T)$ & \multicolumn{7}{c}{$\lfloor \frac{3+w-k}{2} \rfloor$} \\
$T^{+}_{w,b}(z)$ & 1 & 1 & 2 & 4 & 7 & 13 & 25 \\ 
$T^{-}_{w,b}(z)$ & 0 & 1 & 1 & 3 & 5 & 11 & 20 \\ 
\end{tabular}
\caption{Number of independent (anti-)symmetric polynomials and SVMPLs for each $w$.}
\label{table:basis_dimensions}
\end{center}
\end{table}

It is instructive to rewrite the flat space amplitude \eqref{flat_VS} in terms of the proposed ansatz.
Noting that
\bea
{}& \frac{\Gamma(1-S)\Gamma(1-T)\Gamma(1-U)}{\Gamma(1+S)\Gamma(1+T)\Gamma(1+U)}
= \frac{S T}{U}\int d^2z |z|^{-2S-2} |1-z|^{-2T-2}\\
&= \frac{U T}{S}\int d^2z |z|^{-2U-2} |1-z|^{-2T-2}
= \frac{S U}{T}\int d^2z |z|^{-2S-2} |1-z|^{-2U-2}\,,
\eea{Gamma_int}
we can write \eqref{flat_VS} in terms of integrands which are (anti-)symmetric polynomials of degree two
\bea
G^{(0)}_{2,1}(S,T) &= 0\,,& \quad
G^{(0)}_{2,2}(S,T) &= \frac{T^2+S^2}{4}\,,& \quad
G^{(0)}_{2,3}(S,T) &= \frac{T^2-S^2}{4}\,,\\
G^{(0)}_{5,1}(S,T) &= \frac{S T}{2}\,,& \quad
G^{(0)}_{5,2}(S,T) &= 0 \,,& \quad
G^{(0)}_{5,3}(S,T) &= 0\,.
\eea{G0}
Using \eqref{Gamma_int}, we see that this matches the corresponding components of \eqref{flat_VS}
\bea
\frac{S}{U T} \left(G^{(0)}_{2,2}(U,T) + G^{(0)}_{2,3}(U,T) \right)
+ \frac{T}{S U} \left(G^{(0)}_{2,2}(S,U) - G^{(0)}_{2,3}(S,U) \right) &= \frac{S T}{U}\,,\\
\frac{U}{S T} G^{(0)}_{5,1}(S,T) &= \frac{U}{2}\,.
\eea{G0_match}

\subsection{Ambiguities}

It turns out that the above ansatz contains a lot of ambiguities that integrate to 0. 
To give an idea how these ambiguities look like, let us consider the simplest example, which contains only weight 0 and 1 terms.
It is given by (note that $\cL_0(z) = \log |z|^2$ and $\cL_1(z) = \log |1-z|^2$)
\bea
G^\text{amb}_{2,1}(S,T,z) &= G^\text{amb}_{2,2}(S,T,z) = 2(S+T) - S^2 \log |z|^2  - T^2 \log |1-z|^2 \,,\\
G^\text{amb}_{2,3}(S,T,z) &= 0\,, \qquad
G^\text{amb}_{5,j}(S,T,z) = 0\,.
\eea{amb_example}
Using
\beq
I(S,T) = \int d^2z |z|^{-2S-2} |1-z|^{-2T-2} = \frac{\Gamma(-S)\Gamma(-T)\Gamma(1+S+T)}{\Gamma(1+S)\Gamma(1+T)\Gamma(-S-T)}\,,
\eeq
we can replace the logarithms in \eqref{amb_example} by derivatives and check that
\beq
A^\text{amb}_2(S,T)=
\left(2(S+T) + S^2 \partial_S + T^2 \partial_T\right) I(S,T) + (S \leftrightarrow U)  + (T \leftrightarrow U) = 0\,.
\eeq
The presence of ambiguities means that we cannot provide a unique answer for the integrands. However, the AdS amplitude corrections $A^{(1)}(S,T)$ and $A^{(2)}(S,T)$ will be uniquely determined. Below we will display integrands with a choice of ambiguities, and provide the full results with all ambiguities in an ancillary \texttt{Mathematica} file.

\section{First curvature correction}
\label{sec:A1}

Let us now discuss the case $k=1$. Here the ansatz \eqref{G_ansatz_symmetrized} has 166 parameters and 84 ambiguities, so that 82 of the coefficients determine $A^{(1)}(S,T)$.
 By matching the ansatz with the pole structure dictated by the flat space limit \eqref{R_def} of the OPE \eqref{superOPE}, and the known SUGRA term in \eqref{A_SG}, we can completely fix these 82 coefficients. Note that in order to match all inequivalent poles, we consider the $S$-channel poles of the structures $i=1,2,4,5$.

In order to present the result, we make a choice for the 84 ambiguities. It is possible to choose them in such a way that all the $G^{(1)}_{i,j}(S,T,z)$ have homogeneous weight 3. This choice fixes 60 of the ambiguities. 
We then fix the remaining 24 ambiguities arbitrarily and obtain
\begin{align}
{}&G^{(1)}_{2,1}(S,T,z) = \frac{1}{192} \left(111 S^4-31 S^3 T-31 S T^3+111 T^4\right)
   \cL^+_{000}(z)\nonumber\\
&+\frac{1}{192} (S^2-T^2) \left(111 S^2-31 S T+111
   T^2\right) \cL^-_{000}(z)\,,\nonumber\\
{}&G^{(1)}_{2,2}(S,T,z) = \frac{1}{192} \left(-31 S^4+192 S^3 T+247 S^2 T^2+192 S T^3-31
   T^4\right) \cL^+_{000}(z)
+\frac{1}{96} \big(-58 S^4\nonumber\\
&-117 S^3
   T-129 S^2 T^2-117 S T^3-58 T^4\big)
   \cL^+_{001}(z)
+\frac{1}{192} \big(92 S^4-37 S^3 T-37 S^2 T^2-37
   S T^3\nonumber\\
&+92 T^4\big) \cL^+_{010}(z)
+ (S^2-T^2) \bigg(
\frac{-1}{192}  \left(31
   S^2-30 S T+31 T^2\right) \cL^-_{000}(z)
+\frac{1}{48} 
   \big(40 S^2+51 S T\nonumber\\
&+40 T^2\big) 
   \cL^-_{001}(z)
+\frac{1}{96} \left(11 S^2-S T+11 T^2\right)
    \cL^-_{010}(z)\bigg)
-\frac{23}{6}  \left(S^2+T^2\right)
   (S+T)^2\zeta (3)\,,\label{G12_result}\\
{}&G^{(1)}_{2,3}(S,T,z) = (S^2-T^2) \bigg(\frac{1}{192}  \left(142 S^2+61 S T+142 T^2\right) 
   \cL^+_{000}(z)
+\frac{1}{96}  \big(58 S^2+117 S T\nonumber\\
&+58 T^2\big) \cL^+_{001}(z)
-\frac{1}{192}  \left(92
   S^2-37 S T+92 T^2\right)  \cL^+_{010}(z) \bigg)
+\frac{1}{192}
   \big(142 S^4+223 S^3 T+247 S^2 T^2\nonumber\\
&+223 S T^3+142 T^4\big)   \cL^-_{000}(z)
+\frac{1}{96} \left(-80 S^4-102 S^3 T-129 S^2
   T^2-102 S T^3-80 T^4\right) \cL^-_{001}(z)\nonumber\\
&+\frac{1}{192}
   \left(-22 S^4+2 S^3 T-37 S^2 T^2+2 S T^3-22 T^4\right)
   \cL^-_{010}(z)
+\frac{23}{6}  (S^2-T^2) (S+T)^2\zeta (3)\,,\nonumber
\end{align}
as well as
\begin{align}
{}&G^{(1)}_{5,1}(S,T,z) = \frac{1}{48} \big(-S^4+63 S^2 T^2-T^4\big)
   \cL^+_{010}(z)
+\frac{1}{48} \big(-5 S^4+28 S^3 T+5 S^2 T^2+28 S
   T^3\nonumber\\
&-5 T^4\big) \cL^+_{000}(z)
-\frac{1}{8} (S+T)^4
   \cL^+_{001}(z)
(S^2-T^2) \bigg(\frac{-1}{48}  \left(5 S^2-28 S T+5 T^2\right)
    \cL^-_{000}(z)\nonumber\\
&-\frac{1}{8}  (S+T)^2
   \cL^-_{001}(z)
-\frac{1}{48}  (S+T)^2
   \cL^-_{010}(z)\bigg)
-\frac{2}{3}  (S+T)^4\zeta (3)\,,\nonumber\\
{}&G^{(1)}_{5,2}(S,T,z) = \frac{1}{96} \left(S^2+S T+T^2\right) \left(S^2+17 S T+T^2\right)
   \cL^+_{010}(z)
+\frac{1}{96} \big(14 S^4-14 S^3 T\nonumber\\
&-5 S^2 T^2-14 S
   T^3+14 T^4\big) \cL^+_{000}(z)
+\frac{1}{48} \left(3 S^4-2 S^3
   T-28 S^2 T^2-2 S T^3+3 T^4\right) \cL^+_{001}(z)\nonumber\\
&+(S^2-T^2) \bigg(\frac{7}{48}
    \left(S^2-S T+T^2\right)  \cL^-_{000}(z)
+\frac{1}{48}
    \left(3 S^2-8 S T+3 T^2\right) 
   \cL^-_{001}(z)\label{G15_result}\\
&+\frac{1}{96}  \left(S^2+20 S T+T^2\right)
   \cL^-_{010}(z)\bigg)
+\frac{1}{3}  \left(S^2-14 S
   T+T^2\right) (S+T)^2 \zeta (3)\,,\nonumber\\
{}&G^{(1)}_{5,3}(S,T,z) = \frac{S^2 - T^2}{48} \bigg( -  \left(19 S^2+27 S T+19 T^2\right)
   \cL^+_{000}(z)
+  \left(3 S^2+4 S T+3
   T^2\right) \cL^+_{001}(z)\nonumber\\
&+\frac{1}{2}  \left(S^2-18 S
   T+T^2\right) \cL^+_{010}(z) \bigg)
+\frac{1}{96} \left(-38 S^4-54 S^3 T+5
   S^2 T^2-54 S T^3-38 T^4\right) \cL^-_{000}(z)\nonumber\\
&+\frac{1}{48}
   \left(3 S^4-2 S^3 T-20 S^2 T^2-2 S T^3+3 T^4\right)
   \cL^-_{001}(z)
+\frac{1}{96} \big(S^4-16 S^3 T-77 S^2 T^2\nonumber\\
&-16 S   T^3+T^4\big) \cL^-_{010}(z)
+\frac{1}{3}  (S^2-T^2) 
   \left(S^2+9 S T+T^2\right)\zeta (3)\,.
\nonumber
\end{align}
The same result including all ambiguities can be found in an ancillary \texttt{Mathematica} notebook.

\subsection{OPE data}
\label{sec:A1_OPE}

Along with $A^{(1)}(S,T)$ we can fix the following OPE data.
First of all, the leading OPE coefficients for the even spin, $\cZ$-even operators are now determined as
\beq
2 \< f^{(0)}_{1,\delta,\ell} \>
= 3 \< f^{(0)}_{2,\delta,\ell} \>
= \< f^{(0)}_{3,\delta,\ell} \>\,,
\eeq
with $\< f^{(0)}_{3,\delta,\ell} \>$ as given in \eqref{f0_even_results}.
 For the first three odd spin Regge trajectories we find
\bea
\< f^{(0)} \tau^{(2)} \>_{1,\delta,2 \delta-1} ={}&
\frac{r^\text{odd}_1(\delta)}{8} \left(24 \delta ^2-8 \delta +9\right) \,,\\
\< f^{(0)} \tau^{(2)} \>_{1,\delta,2 \delta-3} ={}&
\frac{ r^\text{odd}_2(\delta)}{144} \left(144 \delta ^5+416 \delta ^4-1530 \delta
   ^3+1642 \delta ^2-1167 \delta +306\right)
\,,\\
\< f^{(0)} \tau^{(2)} \>_{1,\delta,2 \delta-5} ={}&
\frac{r^\text{odd}_3(\delta)}{10800} \big(1800 \delta ^8+10640 \delta ^7-27373 \delta ^6-120589
   \delta ^5+383786 \delta ^4\\
&-548839 \delta ^3+543074 \delta   ^2-164055 \delta -78750 \big)
 \,,
\eea{tau2_odd}
and
\begin{align}
\< f^{(2)}_{1,\delta,2 \delta-1} \> ={}&
\frac{r^\text{odd}_1(\delta)}{384} \left(-1792 \delta ^3+2880 \delta ^2-1856 \delta
   +915\right)
+8 \delta ^3 \zeta (3)
r^\text{odd}_1(\delta)\,,\nonumber\\
\< f^{(2)}_{1,\delta,2 \delta-3} \> ={}&
\frac{r^\text{odd}_2(\delta)}{6912}\big(-10752 \delta ^6-46208 \delta ^5+190080 \delta ^4-300046
   \delta ^3+267798 \delta ^2\nonumber\\
&-90153 \delta
   -810\big)
+\frac{4}{3} \left(2 \delta ^3+6 \delta ^2-17
   \delta +6\right) \delta ^3 \zeta (3)
r^\text{odd}_2(\delta)\,,\nonumber\\
\< f^{(2)}_{1,\delta,2 \delta-5} \> ={}&
\frac{r^\text{odd}_3(\delta)}{2592000}\big(-672000 \delta ^9-6721600 \delta ^8+13264256 \delta
   ^7+78009173 \delta ^6-249639187 \delta ^5\nonumber\\
&+352701158 \delta
   ^4-214148773 \delta ^3-174091950 \delta ^2+184200975 \delta
   +40020750\big)\nonumber\\
&+\frac{4}{45} \left(5 \delta ^6+29 \delta
   ^5-58 \delta ^4-373 \delta ^3+722 \delta ^2-97 \delta
   -210\right) \delta ^3 \zeta (3)
r^\text{odd}_3(\delta)\,.
\label{f2_odd}
\end{align}
In particular we can use that the leading odd spin Regge trajectory is non-degenerate to extract
\beq
\tau^{(2)}_{1,\delta,2\delta-1} = \frac{24 \delta^2 - 8 \delta + 9}{32}\,,
\eeq
which agrees with the integrability result \cite{Gromov:2014eha}. For the first three even spin, $\cZ$-odd Regge trajectories we find
\begin{align}
\< f^{(0)} \tau^{(2)} \>_{3,\delta,2 \delta-2} ={}&
\frac{r^\text{even}_1(\delta)}{16 \delta }\big(24 \delta ^2-16 \delta +13 \big)
\,,\nonumber\\
\< f^{(0)} \tau^{(2)} \>_{3,\delta,2 \delta-4} ={}&
\frac{r^\text{even}_2(\delta)}{144} \left(72 \delta ^4+112 \delta ^3-421 \delta ^2+490
   \delta -435\right)
 \,,\nonumber\\
\< f^{(0)} \tau^{(2)} \>_{3,\delta,2 \delta-6} ={}&
\frac{r^\text{even}_3(\delta)}{21600}\big(1800 \delta ^7+6440 \delta ^6-13313 \delta ^5-77327 \delta
   ^4+174881 \delta ^3-291183 \delta ^2\nonumber\\
&+260307 \delta
   +127575\big)
 \,,
\label{tau2_3}
\end{align}
and
\begin{align}
\< f^{(2)}_{3,\delta,2 \delta-2} \> ={}&
-\frac{r^\text{even}_1(\delta)}{768 \delta } \left(1792 \delta ^3-2880 \delta ^2+3584
   \delta -1395\right)
+4 \delta ^2 \zeta (3)r^\text{even}_1(\delta)\,,\nonumber\\
\< f^{(2)}_{3,\delta,2 \delta-4} \> ={}&
\frac{ r^\text{even}_2(\delta)}{6912}\big(-5376 \delta ^5-17728 \delta ^4+54592 \delta ^3-80647
   \delta ^2+93906 \delta -32445\big)\nonumber\\
&+\frac{4}{3} \left(\delta
   ^2+2 \delta -5\right) \delta ^3 \zeta (3)
 r^\text{even}_2(\delta)\,,\nonumber\\
\< f^{(2)}_{3,\delta,2 \delta-6} \> ={}&
\frac{r^\text{even}_3(\delta)}{5184000}\big(-672000 \delta ^8-5377600 \delta ^7+8510656 \delta
   ^6+49491829 \delta ^5-101562397 \delta ^4\nonumber\\
&+136928691 \delta
   ^3-81944109 \delta ^2-147586455 \delta
   +19774125\big)\nonumber\\
&+\frac{2}{45} \left(5 \delta ^5+19 \delta
   ^4-29 \delta ^3-253 \delta ^2+249 \delta +189\right) \delta
   ^3 \zeta (3)
 r^\text{even}_3(\delta)\,.
\label{f2_3}
\end{align}
For the even spin, $\cZ$-even operators we can again only fix a combination, so we show here the result for the leading Regge trajectory and present further results when we can resolve the coefficients independently in section \ref{sec:A2_OPE}
\bea
\< f^{(0)} \tau^{(2)} \>_{1,\delta,2 \delta-2}+
\frac32  \< f^{(0)} \tau^{(2)} \>_{2,\delta,2 \delta-2}={}&
\frac{r^\text{even}_1(\delta)}{16 \delta }\big(24 \delta ^2-16 \delta +1\big)
\,,\\
\< f^{(2)} \>_{1,\delta,2 \delta-2} +
\frac32 \< f^{(2)} \>_{2,\delta,2 \delta-2} ={}&
-\frac{r^\text{even}_1(\delta)}{768 \delta }\big(1792 \delta ^3-2880 \delta ^2+3584
   \delta -1683\big)\\
&+4 \delta ^2 \zeta (3)
r^\text{even}_1(\delta)\,.
\eea{tau2f2_12}
By assuming that there is a unique operator with $(\delta,\ell) = (1,0)$,
we can extract its dimension and again find a match with integrability \cite{Bombardelli:2018bqz}
\beq
\tau^{(2)}_{1,1,0} = \tau^{(2)}_{2,1,0} = \frac{9}{32}\,.
\eeq

\subsection{Low energy expansion}
\label{sec:A1_LEE}

We can directly compute the worldsheet integral in a low energy expansion around $S\sim T\sim 0$ using the method described in \cite{Vanhove:2018elu,Alday:2023jdk}. 
The upshot is that an integral of the form
\beq
I_W(S,T)= \int d^2z |z|^{-2S-2}|1-z|^{-2T-2} {\cal L}_W(z)\,,
\label{I_W}
\eeq
has the low energy expansion
\beq
I_W(S,T)= \text{polar}+\sum_{p,q=0}^\infty (-S)^p (-T)^q \hspace{-15pt} \sum_{W'\in 0^p \shuffle 1^q \shuffle W} \hspace{-15pt} \left( {\cal L}_{0W'}(1)-{\cal L}_{1W'}(1) \right)\,,
\eeq
where the calculation of the polar terms is detailed in \cite{Alday:2023jdk} and $\shuffle$ is the shuffle product. This formula proves that terms of degree $p+q$ in $S$ and $T$ have coefficients which are single-valued multiple zeta values of weight $1+p+q+|W|$, where $|W|$ is the weight of the SVMPL in \eqref{I_W}.
Using this, we get the following result for the low energy expansion
\begin{align}
A^{(1)}_2 (S,T) ={}&
\frac{S^2+3 S T+T^2}{6 U^2}
+\frac{23}{3} S T U \zeta (3)
+\frac{1}{12}  S T U \left(118 S^2+253 S T+118 T^2\right) \zeta (5)\nonumber\\
&+\frac{1}{3} S^2 T^2 \left(6 S^2+77 S T+6 T^2\right) \zeta (3)^2
+\frac{1}{8} S T U \big(80 S^4+412 S^3 T+27 S^2 T^2\nonumber\\
&+412 S T^3+80 T^4\big) \zeta (7)+\ldots\,,\nonumber\\
A^{(1)}_5 (S,T) ={}&
-\frac{3 S^2+7 S T+3 T^2}{12 S T}
-\frac{1}{6} U  \left(41 S^2+105 S T+41 T^2\right)\zeta (3)\label{A1_LEE}\\
&-\frac{1}{12} U 
   \left(194 S^4+671 S^3 T+910 S^2 T^2+671 S T^3+194 T^4\right)\zeta (5)\nonumber\\
&-\frac{1}{6} S T U^2  \left(181 S^2+359 S T+181 T^2\right)\zeta (3)^2
-\frac{1}{16}  U  \big(472 S^6+2080 S^5 T\nonumber\\
&+4793 S^4 T^2+6394 S^3 T^3+4793 S^2 T^4+2080 S T^5+472 T^6\big)\zeta (7)+\ldots\,.
\nonumber
\end{align}
For each of the two results, the leading term originates from supergravity and we used the subleading terms in the supergravity amplitude \eqref{A_SG} to fix a few of the coefficients of the worldsheet ansatz. All other terms in \eqref{A1_LEE} are predictions for parts of the derivative interactions in the low energy effective action. We can immediately compare the terms $\sim \zeta(3)$ in \eqref{A1_LEE} to the subleading terms in the $R^4$ interaction \eqref{A_R4} and see that they match perfectly. 

Once we have fixed the second curvature correction, we will also use the terms $\sim \zeta(5)$ to fully fix the $D^4 R^4$ interaction in section \ref{sec:A2_LEE}.

\subsection{High energy limit}
\label{sec:A1_HE}

We can also compare our results in the fixed-angle high energy limit $|S|,|T|\gg 1$ with $S/T$ fixed to the classical scattering computation of \cite{Alday:2023pzu}. 
We start by noting that the flat space amplitude \eqref{flat_VS} is
\beq
A^{(0)}(S,T) =
\left( \frac{T U}{S} , \frac{S T}{U}, \frac{S U}{T}, \frac{S}{2}, \frac{U}{2}, \frac{T}{2}\right)
\frac{S T U}{U^2}
\int d^2z |z|^{-2S-2} |1-z|^{-2T-2}\,.
\eeq
The high energy limit is dominated by the saddle point $z = \frac{S}{S+T}$ where the integral becomes
\beq
A^{(0)}_\text{HE}(S,T) \propto
\left( \frac{T U}{S} , \frac{S T}{U}, \frac{S U}{T}, \frac{S}{2}, \frac{U}{2}, \frac{T}{2}\right)
\frac{S T U}{U^2} e^{-2 S \log |S|-2 T \log |T|-2 U \log |U|}\,.
\eeq
The high energy limit of the curvature corrections can be obtained in the same way, by evaluating the integral \eqref{Gtot_def} at the saddle point. In this case we find
\beq
A^{(1)}_\text{HE}(S,T) = A^{(0)}_\text{HE}(S,T) S^2 W_3 (z_0)\,,
\eeq
where $z_0 = \frac{S}{S+T}$ and
\bea
W_3 (z_0) ={}&
\mathcal{L}_{000}\left(z_0\right)
-\mathcal{L}_{001}  \left(z_0\right)
-\frac{ 1}{z_0}\mathcal{L}_{010}\left(z_0\right)
-\frac{\left(z_0-1\right){}^2}{z_0^2}\mathcal{L}_{011}\left(z_0\right)\\
&+\frac{\left(z_0-1\right)    }{z_0^2}\mathcal{L}_{101}\left(z_0\right)
+\frac{\left(z_0-1\right){}^2 }{z_0^2}\mathcal{L}_{111}\left(z_0\right)
+2 \zeta (3)\,.
\eea{W3}
This is compatible with the main result of \cite{Alday:2023pzu}, which says that in the limit of large $S$, $T$ and $R$ with $S/T$ and $S/R$ fixed, the amplitude is given by
\beq
A_\text{HE}(S,T) = A^{(0)}_\text{HE}(S,T) e^{-\frac{\alpha'}{R^2}(\mathcal{S}^{(1)} + 2 S F_2 (z_0) \mathcal{S}^{(0)})}\,,
\label{HE_general}
\eeq
where $\mathcal{S}^{(0)}$ and $\mathcal{S}^{(1)}$ are the leading and subleading contributions of the action evaluated on the classical solution for the scattering problem in AdS. They are of weight 1 and 3 respectively. $F_2 (z_0)$ is a function of weight 2 that could not be fixed by the classical computation.
By comparing $W_3 (z_0)$ to the expressions for $\mathcal{S}^{(0)}$ and $\mathcal{S}^{(1)}$ in \cite{Alday:2023pzu}
\beq
S^2 W_3 (z_0) = -\mathcal{S}^{(1)} - 2 S F_2 (z_0) \mathcal{S}^{(0)}\,,
\eeq
we see that
\beq
F_2(z_0) = \frac{1}{4} \left(-  \mathcal{L}_{00}\left(z_0\right)
+\frac{2}{z_0} \mathcal{L}_{01}\left(z_0\right)
+\frac{ z_0-1}{z_0} \mathcal{L}_{11}\left(z_0\right) \right)\,,
\label{F2}
\eeq
which is exactly the same as obtained from the AdS Virasoro-Shapiro amplitude in $AdS_5 \times S^5$ and $AdS_3\times S^3\times M_4$.

\section{Second curvature correction}
\label{sec:A2}

For $k=2$ the ansatz \eqref{G_ansatz_symmetrized} has 1692 parameters. We are not quite able to fix all these parameters, so we make the further assumption that the ansatz has uniform weight, i.e.\ we only include the terms with $w=6$ in \eqref{G_ansatz_symmetrized}.
As we will see, this will allow us to fix all parameters and several non-trivial checks give us confidence that it is the right ansatz.
The uniform weight ansatz still has 950 parameters and 192 ambiguities. Of the remaining 758 parameters, 749 are fixed by matching the poles with the OPE and the supergravity terms \eqref{A_SG}.

To fix the remaining parameters, we impose that the leading odd and even Regge trajectories are non-degenerate, i.e.\ 
\bea
\< f^{(0)} (\tau^{(2)})^2 \>_{1,\delta,2 \delta-1} ={}&
\frac{\< f^{(0)} \tau^{(2)} \>_{1,\delta,2 \delta-1}^2}{\< f^{(0)}  \>_{1,\delta,2 \delta-1}}\,,\\
\< f^{(0)} (\tau^{(2)})^2 \>_{3,\delta,2 \delta-2} ={}&
\frac{\< f^{(0)} \tau^{(2)} \>_{3,\delta,2 \delta-2}^2}{\< f^{(0)}  \>_{3,\delta,2 \delta-2}}\,,\\
\< f^{(0)} (\tau^{(2)})^2 \>_{1,\delta,2 \delta-2}
+\frac32 \< f^{(0)} (\tau^{(2)})^2 \>_{2,\delta,2 \delta-2}
 ={}&
\frac{\< f^{(0)} \tau^{(2)} \>_{1,\delta,2 \delta-2}^2}{\< f^{(0)}  \>_{1,\delta,2 \delta-2}}
+ \frac32 \frac{\< f^{(0)} \tau^{(2)} \>_{2,\delta,2 \delta-2}^2}{\< f^{(0)}  \>_{2,\delta,2 \delta-2}}\,.
\eea{no_mixing}
This fixes 6 parameters.
Next we insert the dimension of the first operator on the leading odd spin Regge trajectory from \cite{Gromov:2014eha}
\beq
\tau^{(4)}_{1,1,1} = \frac{271}{2048}-\frac{9}{8} \zeta (3)\,,
\eeq
which fixes 2 parameters.
Finally, we find for the contribution of $A^{(2)}(S,T)$ to the $R^4$ interaction
\bea
A_2^{(2),\text{R4}}(S,T) &= \frac{\zeta (3)}{18} \left(80 S^2+(221+2 c^{(2)}) S T+80 T^2\right)  \,,\\
A_5^{(2),\text{R4}}(S,T) &= \frac{\zeta (3)}{9}  \left((133+c^{(2)}) S^2+(286+2c^{(2)}) S T+(133+c^{(2)}) T^2\right)\,,
\eea{A2_match_R4}
and comparing to \eqref{A_R4} we fix the final coefficient $c^{(2)}$ to zero.

With all parameters fixed, the worldsheet integrand for the second curvature correction is still quite lengthy and depends on ambiguities, so we only provide it in the attached $\texttt{Mathematica}$ file.

\subsection{OPE data}
\label{sec:A2_OPE}

For the OPE data extracted from the second curvature correction, we mostly present the data for the leading Regge trajectories with $\ell = 2 \delta -1 $ and $\ell = 2 \delta -2 $, and include all data for the subleading Regge trajectories $\ell = 2 \delta -3 $ and $\ell = 2 \delta -4 $ in the attached $\mathtt{Mathematica}$ notebook.

With the second curvature correction fixed, we can resolve the OPE data for the even spin, $\cZ$-even operators that was previously only partially fixed in \eqref{tau2f2_12}
\begin{align}
2\< f^{(0)} \tau^{(2)} \>_{1,\delta,2 \delta-2} =
3 \< f^{(0)} \tau^{(2)} \>_{2,\delta,2 \delta-2} ={}&
\frac{r^\text{even}_1(\delta)}{16 \delta }\big(24 \delta ^2-16 \delta +1\big)
\,,\nonumber\\
2\< f^{(0)} \tau^{(2)} \>_{1,\delta,2 \delta-4} =
3\< f^{(0)} \tau^{(2)} \>_{2,\delta,2 \delta-4} ={}&
\frac{r^\text{even}_2(\delta)}{144} \left(72 \delta ^4+112 \delta ^3-457 \delta ^2+418 \delta
   -255\right)
 \,,\nonumber\\
2\< f^{(0)} \tau^{(2)} \>_{1,\delta,2 \delta-6} =
3\< f^{(0)} \tau^{(2)} \>_{2,\delta,2 \delta-6} ={}&
\frac{ r^\text{even}_3(\delta)}{21600} \big(1800 \delta ^7+6440 \delta ^6-14213 \delta ^5-80747 \delta ^4\nonumber\\
&+180101 \delta ^3-245643 \delta ^2+215487 \delta +93555\big)
\,,
\label{tau2_1}
\end{align}
and
\begin{align}
2\< f^{(2)} \>_{1,\delta,2 \delta-2} ={}&
\frac{r^\text{even}_1(\delta)}{768 \delta }
\big(-1792 \delta ^3+2880 \delta ^2-3584 \delta +3699\big)+4 \delta ^2 \zeta (3) r^\text{even}_1(\delta)\,,\nonumber\\
2\< f^{(2)} \>_{1,\delta,2 \delta-4} ={}&
\frac{r^\text{even}_2(\delta)}{6912}
\big(-5376 \delta ^5-17728 \delta ^4+54592 \delta ^3-73735 \delta ^2+107730 \delta  -67005\big)\nonumber\\
&+\frac{4}{3} \left(\delta ^5+2 \delta ^4-5 \delta ^3\right) \zeta (3) r^\text{even}_2(\delta)
 \,, \nonumber\\
3\< f^{(2)} \>_{2,\delta,2 \delta-2} ={}&
\frac{r^\text{even}_1(\delta)}{768 \delta }
\big(-1792 \delta ^3+2880 \delta ^2-3584 \delta -333 \big)+4 \delta ^2 \zeta (3)r^\text{even}_1(\delta)
\,,\nonumber\\
3\< f^{(2)} \>_{2,\delta,2 \delta-4} ={}&
\frac{ r^\text{even}_2(\delta)}{6912}
\big(-5376 \delta ^5-17728 \delta ^4+54592 \delta ^3-85831 \delta ^2+83538 \delta -6525\big)\nonumber\\
&+\frac{4}{3} \left(\delta ^5+2 \delta ^4-5 \delta ^3\right) \zeta (3)r^\text{even}_2(\delta)
\,.
\label{f2_2}
\end{align}
We also obtain for the leading odd spin Regge trajectory
\bea
\< f^{(0)} (\tau^{(2)})^2 \>_{1,\delta,2 \delta-1} ={}&
\frac{r^\text{odd}_1(\delta)}{256} \left(576 \delta ^4-384 \delta ^3+496 \delta ^2-144 \delta +81\right)\,,\\
\< f^{(0)} \tau^{(4)} + f^{(2)} \tau^{(2)} \>_{1,\delta,2 \delta-1}={}&
\frac{r^\text{odd}_1(\delta)}{12288}\big(-43008 \delta ^5+51200 \delta ^4-59136 \delta ^3+75400 \delta ^2\\
&-20568 \delta +6291\big)+\frac{1}{4} \left(24 \delta ^5-8 \delta ^4-9 \delta ^3\right) \zeta (3) r^\text{odd}_1(\delta)\,,
\eea{A2_OPE_leading_odd}
as well as
\bea
\< f^{(4)}  \>_{1,\delta,2 \delta-1}={}&
\bigg(\frac{1}{5898240}\big(16056320 \delta ^6+26247168 \delta ^5-56258560 \delta ^4-13570560 \delta ^3\\
&+49159040 \delta ^2-137088 \delta +886095\big)
+\frac{1}{192} \big(-1792 \delta ^6+576 \delta ^5\\
&-3392 \delta ^4-3165 \delta ^3\big) \zeta (3)
-12 \delta ^5 \zeta (5)+8 \delta ^6 \zeta (3)^2\bigg) r^\text{odd}_1(\delta)\,.
\eea{f4_leading_odd}
From \eqref{A2_OPE_leading_odd} we can extract the correction to the conformal dimensions 
\beq
\tau^{(4)}_{1,\delta,2\delta-1} = \frac{1}{2048}\big(-1344 \delta ^4+1024 \delta ^3+528 \delta ^2+144 \delta -81\big)-\frac{9 \delta ^3 }{8}\zeta (3)\,,
\eeq
which agrees with the integrability result of \cite{Gromov:2014eha}.

For the leading even spin, $\cZ$-odd trajectory we have
\begin{align}
\< f^{(0)} (\tau^{(2)})^2 \>_{3,\delta,2 \delta-2} ={}&
\frac{r^\text{even}_1(\delta)}{512 \delta } \big(
576 \delta ^4-768 \delta ^3+880 \delta ^2-416 \delta +169 \big)\,,\label{A2_OPE_leading_n=3}\\
\< f^{(0)} \tau^{(4)} + f^{(2)} \tau^{(2)} \>_{3,\delta,2 \delta-2}={}&
\frac{r^\text{even}_1(\delta)}{24576 \delta }
\big(-43008 \delta ^5+65536 \delta ^4-118528 \delta ^3+134792 \delta ^2\nonumber\\
&-61232 \delta +14079\big)+\frac{1}{8} \left(24 \delta ^2-16 \delta -5\right) \delta ^2
   \zeta (3)r^\text{even}_1(\delta)\,,
\nonumber
\end{align}
and
\begin{align}
\< f^{(4)}  \>_{3,\delta,2 \delta-2}={}&
\bigg(
\frac{1}{11796480 \delta }\big(16056320 \delta ^6+26247168 \delta ^5-39055360 \delta ^4-46625280 \delta ^3\nonumber\\
&+104577920 \delta ^2-77058048 \delta +14712975\big)
-\frac{1}{384} \big(1792 \delta ^3-576 \delta ^2+4352 \delta\nonumber\\
& +3069\big) \delta ^2 \zeta (3)
-6 \delta ^4 \zeta (5)
+4 \delta ^5 \zeta(3)^2
\bigg) r^\text{even}_1(\delta)\,.
\label{f4_leading_n=3}
\end{align}
Finally, for the leading even spin, $\cZ$-even trajectory we find
\begin{align}
{}&2\< f^{(0)} (\tau^{(2)})^2 \>_{1,\delta,2 \delta-2}
+  3\< f^{(0)} (\tau^{(2)})^2 \>_{2,\delta,2 \delta-2}=
\frac{r^\text{even}_1(\delta)}{256 \delta }
\big(576 \delta ^4-768 \delta ^3+304 \delta ^2-32 \delta +1 \big) \,,\nonumber\\
&2\< f^{(0)} \tau^{(4)} + f^{(2)} \tau^{(2)} \>_{1,\delta,2 \delta-2}+
3\< f^{(0)} \tau^{(4)} + f^{(2)} \tau^{(2)} \>_{2,\delta,2 \delta-2}
=
\frac{r^\text{even}_1(\delta)}{12288 \delta }
\big(-43008 \delta ^5\label{tau4_leading_n=12}\\
&+65536 \delta ^4-97024 \delta ^3+93320 \delta ^2-29744 \delta +1659\big)
+\frac{1}{4} \left(24 \delta ^2-16 \delta -17\right) \delta ^2
   \zeta (3)r^\text{even}_1(\delta)\,,
\nonumber
\end{align}
and
\begin{align}
{}&2\< f^{(4)} \>_{1,\delta,2 \delta-2}+
3\< f^{(4)}  \>_{2,\delta,2 \delta-2}
=
\bigg(
\frac{1}{5898240 \delta }\big(16056320 \delta ^6
+26247168 \delta ^5-39055360 \delta ^4\nonumber\\
&-41464320 \delta ^3+107342720 \delta ^2
-129773568 \delta +71434575\big)
-\frac{1}{192} \big(1792 \delta ^3-576 \delta ^2+4352 \delta\nonumber\\ &+3357\big) \delta ^2 \zeta (3)
-12 \delta ^4 \zeta (5)
+8 \delta ^5 \zeta(3)^2
\bigg)
r^\text{even}_1(\delta)\,.
\label{f4_leading_n=12}
\end{align}
Assuming uniqueness of the operator with $(\delta,\ell) = (1,0)$, we can extract its subleading correction to the dimension from \eqref{tau4_leading_n=12}
\beq
\tau^{(4)}_{1,1,0} = \tau^{(4)}_{2,1,0} = -\frac{81}{2048}-\frac{9}{8} \zeta (3)\,,
\eeq
and this matches with the result for the $L=1$, $S=1$ operator from
\cite{Bombardelli:2018bqz}.

\subsection{Low energy expansion and $D^4 R^4$}
\label{sec:A2_LEE}

We compute the low energy expansion of $A^{(2)}(S,T)$ and find
\begin{align}
A^{(2)}_2(S,T) ={}&
-\frac{2 S^2-13 S T+2 T^2}{36 U^3}
+\frac{1}{18}   \left(80 S^2+221 S T+80 T^2\right)\zeta (3)\nonumber\\
&+\frac{1}{8}  \left(74 S^4+427 S^3 T+1008 S^2 T^2+427 S
   T^3+74 T^4\right)\zeta (5)\nonumber\\
&-\frac{1}{12}  S T U  \left(342 S^2+605 S T+342 T^2\right)\zeta (3)^2
+\frac{1}{192}   \big(2080 S^6+15984 S^5 T\nonumber\\
&+76386 S^4 T^2-16463
   S^3 T^3+76386 S^2 T^4+15984 S T^5+2080 T^6\big)\zeta (7)+\ldots\nonumber\\
A^{(2)}_5(S,T) ={}&
\frac{U \left(2 S^2-S T+2 T^2\right)}{24 S^2 T^2}
+\frac{1}{9}   \left(133 S^2+286 S T+133 T^2\right)\zeta (3)\label{A2_LEE}\\
&+\frac{1}{16}   \left(1918 S^4+7979
   S^3 T+12078 S^2 T^2+7979 S T^3+1918 T^4\right)\zeta (5)\nonumber\\
&+\frac{1}{36}  U  \left(6290 S^4+26025 S^3 T+39218 S^2 T^2+26025 S T^3+6290
   T^4\right)\zeta (3)^2\nonumber\\
&+\frac{1}{384}   \big(177952 S^6+1319955 S^5 T+3587403 S^4 T^2+4889968 S^3 T^3\nonumber\\
&+3587403 S^2 T^4+1319955 S T^5+177952
   T^6\big)\zeta (7)+\ldots\,.\nonumber
\end{align}
As we have already matched the first two terms to \eqref{A_SG} and \eqref{A_R4} when we fixed our ansatz, the new predictions start with the $\zeta(5)$ term. Recall from \eqref{Mexp} that there are 8 polynomial Mellin amplitudes that can contribute to the $D^4 R^4$ interaction. We can find the coefficients $b_{6,j}$, $b_{5,j}$ and $b_{4,j}$ of these polynomials by computing the corresponding AdS amplitude using \eqref{FlatLimit} and comparing with the $\zeta(5)$ terms in the low energy expansions of $A^{(0)} (S,T)$ \eqref{flat_VS}, $A^{(1)} (S,T)$ \eqref{A1_LEE} and  $A^{(2)} (S,T)$ \eqref{A2_LEE}, giving
\begin{align}
b_{6,1} &= \frac{6147225 \zeta (5)}{2116}\,, 
&b_{6,2} &= -\frac{6054615 \zeta (5)}{1058}\,, 
&b_{6,3} &= \frac{5869395 \zeta (5)}{1058}\,, \label{b_AdS}\\
b_{5,1} &= -\frac{4201155 \zeta (5)}{1088}\,, \ \ 
&b_{5,2} &= -\frac{1869525 \zeta (5)}{1088}\,, \ \ 
&b_{4,1} &= \frac{6070365 \zeta (5)}{8704}\,, \ \ 
b_{4,2} = -\frac{87255 \zeta (5)}{128}\,.\nonumber
\end{align}
We also have two localization constraints, as reviewed in Appendix \ref{app:loc}, that take the form
\begin{align}
b_{2} &= -\frac{1228 b_{4,2}}{119}-\frac{22347 b_{5,1}}{37030}-\frac{26548 b_{5,2}}{18515}+\frac{21937 b_{6,1}}{21560}-\frac{5941 b_{6,2}}{258720}-\frac{1811 b_{6,3}}{2940}
\,,\label{loc1}\\
 b_{4,1} &= -\frac{16 b_{4,2}}{17}-\frac{1651 b_{5,1}}{12696}+\frac{808 b_{5,2}}{1587}-\frac{955 b_{6,1}}{7392}+\frac{27355 b_{6,2}}{88704}+\frac{2567 b_{6,3}}{5544}
 \,.\label{loc2}
\end{align}
We can use the first localization constraint to fix
\beq
b_2 = \frac{25021017 \zeta (5)}{2176}\,.
\eeq
The second localization constraint is also satisfied with this solution, providing a non-trivial consistency check on \eqref{A2_LEE}.

\subsection{High energy limit}
\label{sec:A2_HE}

As discussed in section \ref{sec:A1_HE}, the high energy limit is fixed by \eqref{HE_general} to all orders in $S/R$.
The high energy limit of the second curvature correction should be given by
\beq
A^{(2)}_\text{HE}(S,T) = A^{(0)}_\text{HE}(S,T) \, \frac{1}{2} \left( S^2 W_3 (z_0) \right)^2\,,
\eeq
and this indeed matches with our result.

\section{Conclusions}
\label{sec:conclusions}

In this paper we computed the first two curvature corrections to the AdS Virasoro-Shapiro amplitude in $AdS_4\times \mathbb{CP}^3$ by combining a Borel transform of the superblock expansion, with a worldsheet ansatz in terms of SVMPLs. The first correction was completely fixed by this, and satisfied consistency checks against the high energy limit, the leading odd spin Regge trajectory from integrability, and the $R^4$ correction at finite AdS curvature as computed from analytic bootstrap combined with localization. Our ansatz also fixed the second curvature correction, after using additional inputs from integrability and localization, and also satisfied several consistency checks.

Our result gives predictions for an infinite set of CFT data of massive string operators that can be used to guide future integrability studies.\footnote{The Quantum Spectral Curve, a state-of-the-art integrability method for computing spectra, see e.g. \cite{Gromov:2017blm, Levkovich-Maslyuk:2019awk} for reviews, is already well-established for ABJM theory \cite{Cavaglia:2014exa,Bombardelli:2017vhk,Lee:2017mhh,Lee:2018jvn,Bombardelli:2018bqz,Bianchi:2018scb}. In addition, there are some recent advances in the study of three-point functions \cite{Yang:2021hrl,Jiang:2023cdm,Wu:2024uix} by using integrability methods.} For instance, we predict the dimensions of operators on the leading $\cZ$-odd Regge trajectory, which have not yet been determined from integrability.
We also compute the OPE coefficients of these leading Regge trajectories for the first time, which could be compared against future studies combining integrability with numerical bootstrap, as recently done in the $AdS_5\times S^5$ case \cite{Caron-Huot:2022sdy, Caron-Huot:2024tzr}.

For subleading Regge trajectories we can only provide averaged OPE data and it would be interesting to unmix part of this data by combining it with integrability, as done in \cite{Julius:2023hre,Julius:2024ewf} for the case of $AdS_5 \times S^5$. This could lead to a better understanding of the recently described Regge bridges in ABJM theory \cite{Brizio:2024nso}, which are spin reflected Regge trajectories that intersect with other Regge trajectories. These Regge bridges have also been observed in \cite{Julius:2024ewf}, where the OPE data on subleading Regge trajectories from the $AdS_5 \times S^5$ Virasoro-Shapiro amplitude was unmixed.

One curious result from our study is that the high energy limit does not only match the classical computation of \cite{Alday:2023pzu}, but that also the subleading piece in \eqref{F2} that is not determined by the classical computation is precisely the same as in the case of $AdS_5\times S^5$ and $AdS_3\times S^3\times M_4$.
It would be interesting to understand why this is the case, and if this should hold more generally in other spacetime dimensions. The fact that our approach works at all is also further evidence that the Borel transform is the natural definition of the AdS amplitude. It would be interesting to justify this more rigorously. 

Lastly, in this work we computed the $D^4R^4$ correction at finite AdS curvature for the first time, but only at genus zero, unlike the $R^4$ term which was computed exactly (i.e. including the genus one correction) in \cite{Binder:2019mpb}. Like the $R^4$ correction, the $D^4R^4$ correction is also protected and only receives finite genus corrections (up to genus two in this case), so one might wonder if it should also be possible to compute it exactly. From the finite AdS curvature analytic bootstrap perspective \cite{Binder:2019mpb}, there are 8 coefficients one must fix. Three can be fixed from the known flat space limit, and two more from the localization constraints derived in \cite{Binder:2019mpb,Binder:2018yvd}. This leaves three more coefficients, that could perhaps be fixed from new integrated constraints coming from the squashed sphere or derivatives of the third mass.\footnote{There are three mass deformations in ABJM theory, but only quartic derivatives of two of those masses have been computed in the type IIA limit \cite{Nosaka:2015iiw,Binder:2019mpb}.} The localization expression for the squashed sphere integrated constraint is already known from \cite{Chester:2020jay,Chester:2021gdw}, but it is still necessary to derive the third mass derivative expression in the type IIA limit.

\section*{Acknowledgements}

We thank Fernando Alday, Nicolò Brizio, Andrea Cavaglià, Nikolay Gromov, Roberto Tateo and Arkady Tseytlin for useful discussions.
TH is supported by the STFC grant ST/X000591/1. SMC and DlZ are supported by the UK Engineering and Physical Sciences Research council grant number EP/Z000106/1, and the Royal Society under the grant URF\textbackslash R1\textbackslash 221310.

\appendix

\section{Polynomial Mellin amplitudes}
\label{app:mellin}

In this appendix, we discuss how to compute the basis of polynomial Mellin amplitudes written in \eqref{Mexp}, which are fixed entirely by superconformal symmetry. The degree 2 and 4 Mellin amplitudes were originally computed in \cite{Binder:2019mpb} by imposing crossing symmetry and various superconformal Ward identities. Unlike the case of correlators of half-BPS multiplets considered in \cite{Chester:2018aca}, the polynomials are not fixed by just imposing the superconformal Ward identity for the bottom component correlator. Instead, one in principle needs Ward identities for correlators of all the operators in the multiplet. For $M^2$ and $M^4$, it turns out that the subset of Ward identities derived in \cite{Binder:2019mpb} are sufficient, but they are not for the higher degree polynomials we now consider.

Instead, we will fix these polynomials by demanding that they have an expansion in superconformal blocks, which encodes all the constraints of superconformal symmetry. In particular, we start with the most generic degree 6 polynomial that is crossing symmetric and satisfies the Ward identities of \cite{Binder:2019mpb}
\begin{align}
  {M}_2^{6} &= (s-2)(t-2) \bigg(-\frac{x_{1,0,2}}{64}  s^2  t^2 
  -\frac{x_{1,2,0}}{16}   \left(s^2+s (3 t-10)+(t-10) t+28\right) (s+t-4)^2\nonumber\\
&  -\frac{x_{1,0,1}}{256}  s  t  \left(7 s^2+s (19 t-68)+t (7 t-68)+176\right)\nonumber\\
&  -\frac{x_{1,1,0}}{32}   \left(s^3+s^2 (5 t-14)+s (t (5 t-38)+68)+(t-6) ((t-8) t+20)\right) (s+t-4)\nonumber\\
&  -\frac{x_{1,0,0}}{64}   \Big(s^4+s^3 (7 t-18)+s^2 (t (13 t-82)+124)+s (t (t (7 t-82)+316)-392)\nonumber\\
&  +t (t ((t-18) t+124)-392)+496\Big)
-\frac{x_{1,4,0}}{4}   (s+t-4)^4
-\frac{x_{1,3,0}}{8}   (s+t-6) (s+t-4)^3\nonumber\\
&  -\frac{x_{1,2,1}}{16}  s  t  (s+t-4)^2
 -\frac{x_{1,1,1}}{64} s  t  (3 s+3 t-16) (s+t-4)\bigg)\, ,
\label{M6_ansatz}
\end{align}
where here we show just the second component.
There are a total of 13 unknown coefficients, given by:
\begin{equation}
x_{1,0,0}, x_{1,0,1}, x_{1,0,2}, x_{1,1,0}, x_{1,1,1}, x_{1,2,0}, x_{1,2,1}, x_{1,3,0}, x_{1,4,0}, x_{4,0,0}, x_{4,0,1}, x_{4,0,2}, x_{4,1,0}.
\end{equation}
This is 5 more than the 8 unknowns expected from the flat space counting in Table 3 of \cite{Binder:2019mpb}.

To expand in superblocks, we then convert the polynomial Mellin amplitude into $\overline{D}$-functions
\begin{multline}
  \overline{D}_{r_1,r_2,r_3,r_4}(U, V) \equiv \int \frac{\dd s}{2\pi \ii} \frac{\dd t}{2\pi \ii} U^{\frac{s}{2}} V^{\frac{t}{2}} \Gamma \left(-\frac{s}{2}\right) \Gamma \left(-\frac{t}{2}\right)
  \Gamma \left(\frac{1}{2} (-r_1-r_2+r_3+r_4)-\frac{s}{2}\right)  \\
 \times \Gamma \left(\frac{1}{2} (r_1-r_2-r_3+r_4)-\frac{t}{2}\right) \Gamma \left(\frac{s+t}{2}+r_2\right) \Gamma \left(\frac{s+t}{2}+\frac{1}{2} (r_1+r_2+r_3-r_4)\right)\,,
\end{multline}
as discussed in Appendix A of \cite{Binder:2019mpb}.

The next step is to verify the expansion of the position space Mellin amplitude in terms of the $\mathcal{N}=6$ superconformal blocks. The superconformal blocks appearing in the stress tensor four-point function includes both short and long blocks. In this analysis, we will focus exclusively on the part of the expansion involving the long blocks, as they are the least constrained by superconformal symmetry.

In practice, we focus on the $\log U$ term in the expansion of the long blocks. This term is proportional to the tree-level OPE coefficient of the exchanged operator multiplied by its anomalous dimension, as discussed in \cite{Chester:2018lbz}. Since this $\log U$ term is absent in the short blocks, it allows us to isolate the contribution of the long blocks.

To extract the $\log U$ term, it is convenient to work in the lightcone limit $U \rightarrow 0$. In this limit, the conformal block\footnote{Recall that the superconformal block can be expressed as a linear combination of bosonic conformal blocks $g_{\Delta, \ell}(U, V)$ with specific prefactors and shifted weights, see \eqref{N=6_block}.} can be expanded as a power series in $U$
\begin{equation}
g_{\Delta, \ell}(U, V) = \sum_{n=0}^\infty U^{\frac{\Delta - \ell}{2} + n} g_{\Delta, \ell}^{[n]}(1 - V),
\end{equation}
where the lightcone block $g_{\Delta, \ell}^{[n]}(1 - V)$ is expressed as a double summation \cite{Li:2019cwm}. Since $g_{\Delta, \ell}^{[n]}(1 - V)$ is independent of $U$, the $\log U$ term can only arise from the expansion of the anomalous dimension $\Delta$ in the exponent $U^{\frac{\Delta - \ell}{2} + n}$ around its generalised free theory value.

On the other hand, the expansion of the $\overline{D}$-functions is also known and is given by \cite{Dolan:2000ut}
\begin{align}
{}&    \overline{D}_{r_1, r_2, r_3, r_4}(U, V) \bigg|_{\log U} = 
    - e^{\frac{1}{2} i \pi (r_1 + r_2 - r_3 - r_4)}
    \sum_{m, n = 0}^\infty U^m (1 - V)^n \\
&    \times
    \frac{\Gamma(r_1 + m) \Gamma(r_2 + m + n) \Gamma\left(\frac{1}{2}(r_1 + r_2 - r_3 + r_4) + m\right) \Gamma\left(\frac{1}{2}(r_1 + r_2 + r_3 - r_4) + m + n\right)}{\Gamma(m + 1) \Gamma(n + 1) \Gamma(r_1 + r_2 + 2m + n) \Gamma\left(\frac{1}{2}(r_1 + r_2 - r_3 - r_4 + 2) + m\right)} \, .\nonumber
\end{align}
Using this expression, we can straightforwardly expand the position-space Mellin amplitude. By comparing the position-space expansion with the $\mathcal{N}=6$ superconformal block expansion in the small-$U$ region, and by matching sufficiently many terms, we find the following five constraints on the coefficients:
\begin{equation}
  \begin{aligned}
    x_{1,4,0} = & \frac{149}{280} x_{1,0,0}-\frac{169}{210} x_{1,0,2}+\frac{263}{105} x_{1,1,0}+\frac{1489}{840} x_{1,1,1}+\frac{93}{140} x_{1,2,0}+\frac{17}{10} x_{1,2,1}\\
    & \qquad -\frac{89}{35} x_{1,3,0}-\frac{1607 }{3360} x_{1,0,1},\\
    x_{4,0,0} = & -\frac{242602 x_{1,0,0}}{8085}+\frac{3124981 x_{1,0,1}}{97020}+\frac{232810 x_{1,0,2}}{4851}-\frac{35132}{385} x_{1,2,1}+\frac{26136}{245} x_{1,3,0} \\
    & \qquad -\frac{461936 x_{1,2,0}}{8085}-\frac{3550322 x_{1,1,0}}{24255}-\frac{2293169 x_{1,1,1}}{24255}, \\
    x_{4,0,1} = & -\frac{21179 x_{1,0,0}}{2695}+\frac{213137 x_{1,0,1}}{32340}+\frac{75706 x_{1,0,2}}{8085}-\frac{7268}{385} x_{1,2,1}+\frac{4192}{245} x_{1,3,0}\\
    & \qquad -\frac{41196 x_{1,2,0}}{2695}-\frac{276004 x_{1,1,0}}{8085}-\frac{160429 x_{1,1,1}}{8085},\\
    x_{4,0,2} = & \frac{300871 x_{1,0,0}}{10780}+\frac{317033 x_{1,1,0}}{2695}+\frac{831727 x_{1,1,1}}{10780}+\frac{121431 x_{1,2,0}}{2695}+\frac{28473}{385} x_{1,2,1} \\
    & \qquad -\frac{19472}{245} x_{1,3,0}-\frac{185359 x_{1,0,2}}{5390}-\frac{871771 x_{1,0,1}}{43120},\\
    x_{4,1,0} = & \frac{146317 x_{1,0,0}}{1617}+\frac{4181791 x_{1,1,0}}{9702}+\frac{27275167 x_{1,1,1}}{97020}+\frac{1333642 x_{1,2,0}}{8085} \\
    & \qquad +\frac{104201}{385} x_{1,2,1}-\frac{76626}{245} x_{1,3,0}-\frac{3342523 x_{1,0,2}}{24255}-\frac{6987613 x_{1,0,1}}{77616}\, .
  \end{aligned}
\end{equation}
After imposing these five constraints, our ${M}_6$ amplitude now depends on only \textit{eight} unknowns:
\begin{equation}
x_{1,0,0}, \quad x_{1,0,1}, \quad x_{1,0,2}, \quad x_{1,1,0}, \quad x_{1,1,1}, \quad x_{1,2,0}, \quad x_{1,2,1}, \quad x_{1,3,0}.
\end{equation}
These unknowns correspond to the polynomials of various degrees: 
\begin{itemize}
\item 3 polynomials of degree 6,
\item 2 polynomials of degree 5,
\item 2 polynomials of degree 4,
\item 1 polynomial of degree 2.
\end{itemize}
This matches the solutions listed in Table 3 of \cite{Binder:2019mpb}. Note that the $f_3$ terms have one less degree than indicated in the table, as they vanish in the flat-space limit. For instance, the degree-7 $f_3$ term in the table corresponds to a degree-6 polynomial in our case. 

In the main text, we will use the $i = 2, 5$ components for all possible polynomials of each degree up to 6. Denoting these polynomials by $M^{\mathtt{degree}, \mathtt{k}}_i$, where the lower index $i$ labels the R-symmetry basis and the superscript $\mathtt{k}$ labels the independent polynomials, we find\footnote{Recall that $s+t+u=4$.} \footnote{We thank Daniele Pavarini for pointing out a typo in this equation.}
\begin{itemize}
   \item Degree 6:
   \begin{equation}
      \begin{aligned}
         M^{6,1}_2 = & -\frac{(s-2) (t-2)}{26880}\Big(26916 s^2 t^2 + 17244 (s^3 t+ s t^3)+ 3996(s^4+t^4) \\
         &  -64776  (s^3+t^3) -206088  (s^2 t+ s t^2) +395376 (s^2+t^2) +819312 s t \\
         &  -1080096 (s+t) +1123776  \Big), \\
         M^{6,1}_5 = &  -\frac{(s+t-2)}{517440} \Big(76923 (s^5+t^5) + 265188(s^4 t + s t ^4) + 344190(s^3 t^2 + s^2 t^3) \\
         &  -2305422  s^2 t^2 - 1591338 (s^3 t+ s t^3)- 514878(s^4+t^4) \\
         &  +1519686 (s^3+t^3) +4773258  (s^2 t+ s t^2) -2820588  (s^2+t^2) - 5837136 s t \\
         & +2496192 (s+t) -1312384 \Big), \\
         M^{6,2}_2 = & \frac{(s-2) (t-2)}{26880} \Big(17289 s^2 t^2 + 12121 (s^3 t+ s t^3)+ 3214(s^4+t^4) \\
         &  -51424  (s^3+t^3) -147132  (s^2 t+ s t^2) +308544 (s^2+t^2) +598608 s t \\
         &  -822784 (s+t) +822784  \Big), \\
         M^{6,2}_5 = & \frac{(s+t-2)}{12418560} \Big(1484868 (s^5+t^5) + 5318313(s^4 t + s t ^4) + 8249010(s^3 t^2 + s^2 t^3) \\
         &  -43506792  s^2 t^2 - 27746208 (s^3 t+ s t^3)- 9032898(s^4+t^4) \\
         & +26161296 (s^3+t^3) +79787568  (s^2 t+ s t^2) -54349488  (s^2+t^2) - 103063536 s t \\
         & +49854192 (s+t) -12232064 \Big), \\
         M^{6,3}_2 = & \frac{(s-2) (t-2) \left(5408 (s+t-4)^4-420 s^2 t^2\right)}{26880}\, ,\\
         M^{6,3}_5 = & \frac{(s+t-2)}{3104640} \Big(624624 (s^5+t^5) + 2210439(s^4 t + s t ^4) + 3220140(s^3 t^2 + s^2 t^3) \\
         &  -18362736  s^2 t^2 - 12124224 (s^3 t+ s t^3)- 3921414(s^4+t^4) \\
         & +11104968 (s^3+t^3) +34792044  (s^2 t+ s t^2) -21347064  (s^2+t^2) - 43540128 s t \\
         & +18773184 (s+t) -5491456 \Big), \\
      \end{aligned}
   \end{equation}
   \item Degree 5:
      \begin{equation}
      \begin{aligned}
         M^{5,1}_2 = & -\frac{1}{8464}(s-2) (t-2)\Big(232(s^3+t^3) + 639 (s^2 t+ s t^2) -3248(s^2+t^2) -6390 s t \\
         & +15776(s+t) -32072  \Big), \\
         M^{5,1}_5 = & -\frac{1}{296240} (s + t-2)\Big(20370 s^2 t^2 + 6125 (s^3 t+ s t^3) -36960  (s^3+t^3) \\
         &  -129710  (s^2 t+ s t^2) +163180 (s^2+t^2) +234120 s t -408864 (s+t) +721152  \Big), \\
         M^{5,2}_2 = & \frac{1}{4232} (s-2) (t-2) \Big(607(s^3+t^3) + 1323 (s^2 t+ s t^2) -8498(s^2+t^2) -14288 s t \\
         &  +39160(s+t) -60144  \Big), \\
         M^{5,2}_5 = & \frac{1}{888720} (s + t-2)\Big(173250 s^2 t^2 + 22890 (s^3 t+ s t^3) -645330  (s^3+t^3) \\
&-1642620  (s^2 t+ s t^2) 
           +3071400 (s^2+t^2) +3649560 s t -3692760 (s+t) \\
&+1796800  \Big)\, . \\
      \end{aligned}
   \end{equation}
   \item Degree 4:
   \begin{equation}
      \begin{aligned}
         M^{4,1}_2 ={}& \frac{1}{35} (s-2) (t-2) \left(35 st-100s-100 t+288 \right), \\   
             M^{4,1}_5 ={}& \frac{1}{70}(s+t-2)\left(224-324(s+t) + (90 s^2+140 s t+90 t^2) + 35 (s^2 t +s t^2) \right), \\
      M^{4,2}_2 ={}& \frac{1}{34}(s-2) (t-2) (17 s^2+49 s t-174 s+17 t^2-174 t+756), \\
     M^{4,2}_5 ={}&\frac{1}{510}(s + t-2)\Big(255 (s^3+t^3)+750 (s^2 t+s t^2)-840 (s^2+t^2) -1080 s t\\
&+6072(s+ t)-15296\Big)\, .  \\
      \end{aligned}
   \end{equation}
   \item Degree 2:
   \begin{equation}
      \begin{aligned}
         M^{2,1}_2 & = (s-2)(t-2), \qquad
         {M}^{2,1}_5 & = \left(u-\frac{4}{3}\right)(u-2)\, . 
      \end{aligned}
   \end{equation}
\end{itemize}
The degree-two polynomials and the first degree-four polynomials were already given in \eqref{R4Mel}.
All polynomials can also be found in the $\texttt{Mathematica}$ file.

\section{OPE poles of the AdS amplitude}
\label{app:ads_poles}

In this appendix we apply the Borel transform \eqref{FlatLimit} to the Mellin amplitude corresponding to a bosonic conformal block
\beq
M_{\tau,\ell}(s,t) = \sum_{m=0}^\infty
\frac{\mathcal{Q}_{\tau,\ell,m}(t-2)}{s-\tau-2m}\,,
\label{Mtauell}
\eeq
where the numerators are Mack polynomials
\begin{align}
\mathcal{Q}_{\tau,\ell, m}(t) = \frac{\Gamma(\ell+\frac12)}{4^{\tau+\ell} \sqrt{\pi} \Gamma(\ell + 1)}  K(1,\tau,\ell,m,3) Q_{\ell, m}^{\tau, d=3}(t) \,, 
\label{curlyQ}
\end{align}
with $Q_{\ell, m}^{\tau, d}(t)$ as in \cite{Alday:2022uxp} and
\beq
K(\Delta,\tau, \ell, m,d)  = - \frac{2 (\ell+\tau -1)_\ell \Gamma (2 \ell+\tau )}{ 2^\ell \Gamma \left(\ell+\frac{\tau}{2} )\right)^4 \Gamma (m+1) \Gamma \left(\Delta -\frac{\tau }{2}-m\right)^2 \left(\ell+\tau-\frac{d}{2} +1\right)_m}.
\eeq
The transform \eqref{FlatLimit} of \eqref{Mtauell} is (where $\beta \equiv 4/3$ is the shift introduced in \eqref{FlatLimit})
\beq
A_{\tau,\ell}(S,T) = 
\frac{\sqrt{\pi}}{4 \sqrt{\nu}}
\int_{\kappa-i \infty}^{\kappa+i \infty} \frac{d \alpha}{2 \pi i} e^{\alpha} \alpha^{-\frac12} 
\sum_{m=0}^\infty
\frac{\mathcal{Q}_{\tau,\ell,m}(\frac{2 \sqrt{\nu} T}{\alpha} + 2 -2\beta)}{\frac{2 \sqrt{\nu} S}{\alpha}+\beta-\tau-2m}\,.
\eeq
We exchange the integral and summation and for each $m$ pick the pole at
\beq
\alpha = \alpha_* \equiv \frac{2 \sqrt{\nu}S}{\tau+2m-\beta}\,.
\eeq
We ignore any other poles in $\alpha$, as we are only interested in $S$-channel poles.
Next we have to sum over $m$. The Mack polynomial for large $\tau$ has its maximum at $m \sim \tau^2$, so we replace the sum over $m=x \tau^2$ by an integral over $x$
\beq
A_{\tau,\ell}(S,T)\Big|_{S\text{-poles}} = -\frac{\sqrt{\pi}}{4 \sqrt{\nu}}
 \frac{\tau^2}{2\sqrt{\nu} S}   \int_0^\infty dx \  e^{\alpha_*} \alpha_*^\frac32
 \mathcal{Q}_{\tau,\ell,x \tau^2}\left(\frac{2 \sqrt{\nu} T}{\alpha_*} + 2 -2\beta\right)\,.
\eeq
We can now expand the integrand at large $\nu$, using that $\tau \sim \nu^{1/4}$.
The answer has the form
\beq
 - \frac{\sqrt{\pi}}{4 \sqrt{\nu}} \frac{\tau^2}{2\sqrt{\nu} S} e^{\alpha_*} \alpha_*^\frac32 
\mathcal{Q}_{\tau,\ell,x \tau^2}\left(\frac{2 \sqrt{\nu} T}{\alpha_*} + 2 -2\beta\right)
= \frac{e^{-\frac{1}{4x} + \frac{\sqrt{\nu}S}{\tau^2 x}}}{x^2} 
\sum_{i=0}^\infty \frac{1}{\nu^{\frac{5}{8}+\frac{i}{4}}} P^{(i)} \left(\tfrac1x\right)\,,
\eeq
where $P^{(i)}\left(\tfrac1x\right)$ are polynomials in $\frac1x$ with increasing degree (which also depend on $\tau,\ell,S,T,\nu$).
We can now do the integrals in $x$ using
\beq
\int_0^\infty dx\, \frac{e^{-\frac{1}{4x} + \frac{\sqrt{\nu}S}{\tau^2 x}}}{x^2} 
= - \frac{\frac{\tau^2}{\sqrt{\nu}}}{S- \frac{\tau^2}{4\sqrt{\nu}}}
= - \frac{4\delta}{S-\delta} + O\left( \nu^{-\frac14}  \right)\,.
\label{x_integral}
\eeq
The effect of the polynomials $P^{(i)} \left(\tfrac1x\right)$ is that each additional power of $\frac1x$ increases the order of the pole by one, as $\frac1x$ can be replaced by $\frac{\tau^2}{\sqrt{\nu}}\partial_S$ acting on both sides of \eqref{x_integral}.
Next we do the $x$ integral and get the formula \eqref{R_def} used in the main text.

\section{Basis of single-valued multiple polylogarithms}
\label{app:SVMPLs}

Below we specify the basis of SVMPLs used in the ansatz \eqref{G_ansatz_symmetrized}
\begin{align}
T^+_{0}(z) ={}& \left(1  \right)\,, \qquad
T^+_{1}(z) = \left(\cL^+_{0}(z)  \right)\,, \qquad
T^-_{1}(z) = \left(\cL^-_{0}(z) \right)\,,\nonumber\\
T^+_{2}(z) ={}& \left(\cL^+_{00}(z), \cL^+_{01}(z)  \right)\,, \qquad
T^-_{2}(z) = \left(\cL^-_{00}(z)\right)\,,\nonumber\\
T^+_{3}(z) ={}& \left(\cL^+_{000}(z), \cL^+_{001}(z), \cL^+_{010}(z), \zeta(3) \right)\,,\nonumber\\
T^-_{3}(z) ={}& \left(\cL^-_{000}(z), \cL^-_{001}(z), \cL^-_{010}(z)  \right)\,,\nonumber\\
T^+_{4}(z) ={}& \left(\cL^+_{0000}(z),\cL^+_{0001}(z),\cL^+_{0010}(z),\cL^+_{0011}(z),\cL^+_{0101}(z),\cL^+_{0110}(z),\zeta (3) \cL^+_0(z) \right)\,,\nonumber\\
T^-_{4}(z) ={}& \left(\cL^-_{0000}(z),\cL^-_{0001}(z),\cL^-_{0010}(z),\cL^-_{0110}(z),\zeta (3) \cL^-_0(z) \right)\,,\nonumber\\
T^+_{5}(z) ={}& \big(\cL^+_{00000}(z),\cL^+_{00001}(z),\cL^+_{00010}(z),\cL^+_{00011}(z),\cL^+_{00100}(z),\cL^+_{00101}(z),\nonumber\\
&\ \,\cL^+_{00110}(z),\cL^+_{01001}(z),\cL^+_{01010}(z),\cL^+_{01110}(z),\zeta (3) \cL^+_{00}(z),\zeta (3) \cL^+_{01}(z),\zeta (5) \big)\,,\label{MPL_list}\\
T^-_{5}(z) ={}& \big(\cL^-_{00000}(z),\cL^-_{00001}(z),\cL^-_{00010}(z),\cL^-_{00011}(z),\cL^-_{00100}(z),\cL^-_{00101}(z),\nonumber\\
&\ \,\cL^-_{00110}(z),\cL^-_{01001}(z),\cL^-_{01010}(z),\cL^-_{01110}(z),\zeta (3) \cL^-_{00}(z) \big)\,,\nonumber\\
T^+_{6}(z) ={}&  \big(
\mathcal{L}^+_{000000}(z),
\mathcal{L}^+_{000001}(z),
\mathcal{L}^+_{000010}(z),
\mathcal{L}^+_{000011}(z),
\mathcal{L}^+_{000100}(z),
\mathcal{L}^+_{000101}(z),
\mathcal{L}^+_{000110}(z),\nonumber\\&
\mathcal{L}^+_{000111}(z),
\mathcal{L}^+_{001001}(z),
\mathcal{L}^+_{001010}(z),
\mathcal{L}^+_{001011}(z),
\mathcal{L}^+_{001100}(z),
\mathcal{L}^+_{001101}(z),
\mathcal{L}^+_{001110}(z),\nonumber\\&
\mathcal{L}^+_{010001}(z),
\mathcal{L}^+_{010010}(z),
\mathcal{L}^+_{010101}(z),
\mathcal{L}^+_{010110}(z),
\mathcal{L}^+_{011001}(z),
\mathcal{L}^+_{011110}(z),\nonumber\\&
\zeta(3)\mathcal{L}^+_{000}(z),
\zeta(3)\mathcal{L}^+_{001}(z),
\zeta(3)\mathcal{L}^+_{010}(z),
\zeta(5)\mathcal{L}^+_{0}(z),
\zeta(3)^2\big)\,,\nonumber\\
T^-_{6}(z) ={}& \big(
\mathcal{L}^-_{000000}(z),
\mathcal{L}^-_{000001}(z),
\mathcal{L}^-_{000010}(z),
\mathcal{L}^-_{000011}(z),
\mathcal{L}^-_{000100}(z),
\mathcal{L}^-_{000101}(z),
\mathcal{L}^-_{000110}(z),\nonumber\\&
\mathcal{L}^-_{001001}(z),
\mathcal{L}^-_{001010}(z),
\mathcal{L}^-_{001100}(z),
\mathcal{L}^-_{001101}(z),
\mathcal{L}^-_{001110}(z),
\mathcal{L}^-_{010001}(z),
\mathcal{L}^-_{010010}(z),\nonumber\\&
\mathcal{L}^-_{010110}(z),
\mathcal{L}^-_{011110}(z),
\zeta(3)\mathcal{L}^-_{000}(z),
\zeta(3)\mathcal{L}^-_{001}(z),
\zeta(3)\mathcal{L}^-_{010}(z),
\zeta(5)\mathcal{L}^-_{0}(z)\big)\,.\nonumber
\end{align}

\section{Localization constraints}
\label{app:loc}

In this appendix we will discuss how the localization constraints of \cite{Binder:2018yvd,Binder:2019mpb} can be used to constrain the $D^4R^4$ term in the low energy expansion. The two localization constraints are
\es{DerSimpOneMassFinal}{
\frac{1}{c_T^2} \frac{\partial \log Z}{\partial m_+^4} &= \frac{\pi^4 }{2^{13}} ( 2 \lambda_{2, 0, {\bf 84}}^2-4) \,,\\
\frac{1}{c_T^2}\frac{\partial \log Z}{\partial m_+^2 \partial m_-^2} &= \frac{\pi^2}{2^{11}}I_{+-}[\mathcal{S}^i].
 }
where the second constraint is given by the integral of the Mellin amplitude:
\begin{equation}\label{stInt}
\begin{split}
I_{+-}[{\cal S}^i] & = \int \frac{ds\ dt}{(4\pi i)^2} \frac{2\sqrt{\pi}}{(2-t)(s+t-2)}M^1(s,t) \\
&\times \Gamma \left[1-\frac{s}{2}\right] \Gamma \left[\frac{s+1}{2}\right] \Gamma \left[1-\frac{t}{2}\right] \Gamma \left[\frac{t-1}{2}\right] \Gamma \left[\frac {s+t-2}{2}\right] \Gamma \left[\frac{3-s-t}2\right]\,.
\end{split}\end{equation}
The OPE coefficient $\lambda_{2, 0, {\bf 84}}^2$ can be extracted by converting Mellin amplitudes to position space as described in Appendix \ref{app:mellin}, and then expanding for the lowest twist 2 block in the $\bf{84}$ irrep of the R-symmetry.

The LHS of these constraints were computed using localization in \cite{Nosaka:2015iiw,Binder:2019mpb} using the Fermi gas formalism developed originally in \cite{Marino:2011eh}, and at large $c_T$ and large $\nu$ take the form
 \es{largeLam}{
 \frac{1}{c_T^2}\frac{\partial^4\log Z}{\partial m_\pm^4}&=\left[\frac{3\pi^2}{64}+\frac{9\zeta(3)\pi^2}{256}\frac{1}{\nu^{\frac32}}+\frac{27 \zeta(3)^2\pi^2}{1024 }\frac{1}{\nu^3}+O(\nu^{-\frac92})\right]\frac{1}{c_T}+O(c_T^{-2})\,,\\
  \frac{1}{c_T^2}\frac{\partial^4\log Z}{\partial m_+^2\partial m_-^2}&=\left[-\frac{\pi^2}{64}-\frac{3\zeta(3)\pi^2}{256}\frac{1}{\nu^{\frac32}}-\frac{9 \zeta (3)^2\pi^2}{1024  }\frac{1}{\nu^3}+O(\nu^{-\frac92})\right]\frac{1}{c_T}+O(c_T^{-2})\,.
}
In particular, they do not contain a $\nu^{-5/2}$ term corresponding to $D^4R^4$. We can then impose these two constraints on the eight coefficients appearing in the $\nu^{-5/2}$ term in \eqref{Mexp} to get the constraints \eqref{loc1} and \eqref{loc2}.

\bibliographystyle{JHEP}
\bibliography{abjm_vs}

@article{Chester:2024wnb,
	archiveprefix = {arXiv},
	author = {Chester, Shai M. and Zhong, De-liang},
	eprint = {2412.06429},
	month = {12},
	primaryclass = {hep-th},
	title = {{The AdS$_3\times $S$^3$ Virasoro-Shapiro amplitude with RR flux}},
	year = {2024}}

@article{Alday:2024rjs,
	archiveprefix = {arXiv},
	author = {Alday, Luis F. and Giribet, Gaston and Hansen, Tobias},
	eprint = {2412.05246},
	month = {12},
	primaryclass = {hep-th},
	title = {{On the $AdS_3$ Virasoro-Shapiro Amplitude}},
	year = {2024}}

@article{Brizio:2024nso,
	archiveprefix = {arXiv},
	author = {Brizio, Nicol\`o and Cavagli\`a, Andrea and Tateo, Roberto and Tripodi, Valerio},
	eprint = {2410.08927},
	month = {10},
	primaryclass = {hep-th},
	title = {{Regge trajectories and bridges between them in integrable AdS/CFT}},
	year = {2024}}

@article{Julius:2023hre,
	archiveprefix = {arXiv},
	author = {Julius, Julius and Sokolova, Nika},
	doi = {10.1007/JHEP03(2024)090},
	eprint = {2310.06041},
	journal = {JHEP},
	pages = {090},
	primaryclass = {hep-th},
	title = {{Conformal field theory-data analysis for $\mathcal{N}$ = 4 Super-Yang-Mills at strong coupling}},
	volume = {03},
	year = {2024}}

@article{Julius:2024ewf,
	archiveprefix = {arXiv},
	author = {Julius, Julius and Sokolova, Nika Sergeevna},
	eprint = {2409.07529},
	month = {9},
	primaryclass = {hep-th},
	title = {{Unmixing sub-leading Regge trajectories of $\mathcal{N} = 4$ Super-Yang-Mills}},
	year = {2024}}

@article{Marino:2011eh,
	archiveprefix = {arXiv},
	author = {Marino, Marcos and Putrov, Pavel},
	doi = {10.1088/1742-5468/2012/03/P03001},
	eprint = {1110.4066},
	journal = {J. Stat. Mech.},
	pages = {P03001},
	primaryclass = {hep-th},
	title = {{ABJM theory as a Fermi gas}},
	volume = {1203},
	year = {2012}}

@article{Nosaka:2015iiw,
	archiveprefix = {arXiv},
	author = {Nosaka, Tomoki},
	doi = {10.1007/JHEP03(2016)059},
	eprint = {1512.02862},
	journal = {JHEP},
	pages = {059},
	primaryclass = {hep-th},
	reportnumber = {YITP-15-105},
	title = {{Instanton effects in ABJM theory with general R-charge assignments}},
	volume = {03},
	year = {2016}}

@article{Chester:2020jay,
	archiveprefix = {arXiv},
	author = {Chester, Shai M. and Kalloor, Rohit R. and Sharon, Adar},
	doi = {10.1007/JHEP07(2020)041},
	eprint = {2004.13603},
	journal = {JHEP},
	pages = {041},
	primaryclass = {hep-th},
	title = {{3d $ \mathcal{N} $ = 4 OPE coefficients from Fermi gas}},
	volume = {07},
	year = {2020}}

@article{Chester:2021gdw,
	archiveprefix = {arXiv},
	author = {Chester, Shai M. and Kalloor, Rohit R. and Sharon, Adar},
	doi = {10.1007/JHEP04(2021)244},
	eprint = {2102.05643},
	journal = {JHEP},
	pages = {244},
	primaryclass = {hep-th},
	title = {{Squashing, Mass, and Holography for 3d Sphere Free Energy}},
	volume = {04},
	year = {2021}}

@article{Binder:2018yvd,
	archiveprefix = {arXiv},
	author = {Binder, Damon J. and Chester, Shai M. and Pufu, Silviu S.},
	doi = {10.1007/JHEP04(2020)052},
	eprint = {1808.10554},
	journal = {JHEP},
	pages = {052},
	primaryclass = {hep-th},
	reportnumber = {PUPT-2570},
	title = {{Absence of $D^4 R^4$ in M-Theory From ABJM}},
	volume = {04},
	year = {2020}}

@article{Caron-Huot:2024tzr,
	archiveprefix = {arXiv},
	author = {Caron-Huot, Simon and Coronado, Frank and Zahraee, Zahra},
	eprint = {2412.00249},
	month = {11},
	primaryclass = {hep-th},
	title = {{Bootstrapping $\mathcal{N} = 4$ sYM correlators using Integrability and Localization}},
	year = {2024}}

@article{Aharony:2008ug,
	archiveprefix = {arXiv},
	author = {Aharony, Ofer and Bergman, Oren and Jafferis, Daniel Louis and Maldacena, Juan},
	doi = {10.1088/1126-6708/2008/10/091},
	eprint = {0806.1218},
	journal = {JHEP},
	pages = {091},
	primaryclass = {hep-th},
	reportnumber = {WIS-12-08-JUN-DPP},
	title = {{N=6 superconformal Chern-Simons-matter theories, M2-branes and their gravity duals}},
	volume = {10},
	year = {2008}}

@article{Alday:2022xwz,
	archiveprefix = {arXiv},
	author = {Alday, Luis F. and Hansen, Tobias and Silva, Joao A.},
	doi = {10.1007/JHEP12(2022)010},
	eprint = {2209.06223},
	journal = {JHEP},
	pages = {010},
	primaryclass = {hep-th},
	title = {{AdS Virasoro-Shapiro from single-valued periods}},
	volume = {12},
	year = {2022}}

@article{Alday:2024ksp,
	archiveprefix = {arXiv},
	author = {Alday, Luis F. and Hansen, Tobias},
	doi = {10.1007/JHEP08(2024)108},
	eprint = {2404.16084},
	journal = {JHEP},
	pages = {108},
	primaryclass = {hep-th},
	title = {{Single-valuedness of the AdS Veneziano amplitude}},
	volume = {08},
	year = {2024}}

@article{Alday:2024yax,
	archiveprefix = {arXiv},
	author = {Alday, Luis F. and Chester, Shai M. and Hansen, Tobias and Zhong, De-liang},
	doi = {10.1007/JHEP05(2024)322},
	eprint = {2403.13877},
	journal = {JHEP},
	pages = {322},
	primaryclass = {hep-th},
	title = {{The AdS Veneziano amplitude at small curvature}},
	volume = {05},
	year = {2024}}

@article{Penedones:2010ue,
	archiveprefix = {arXiv},
	author = {Penedones, Joao},
	doi = {10.1007/JHEP03(2011)025},
	eprint = {1011.1485},
	journal = {JHEP},
	pages = {025},
	primaryclass = {hep-th},
	title = {{Writing CFT correlation functions as AdS scattering amplitudes}},
	volume = {03},
	year = {2011}}

@article{Berkovits:2000fe,
	archiveprefix = {arXiv},
	author = {Berkovits, Nathan},
	doi = {10.1088/1126-6708/2000/04/018},
	eprint = {hep-th/0001035},
	journal = {JHEP},
	pages = {018},
	reportnumber = {IFT-P-005-2000},
	title = {{Super Poincare covariant quantization of the superstring}},
	volume = {04},
	year = {2000}}

@article{Maldacena:1997re,
	archiveprefix = {arXiv},
	author = {Maldacena, Juan Martin},
	doi = {10.4310/ATMP.1998.v2.n2.a1},
	eprint = {hep-th/9711200},
	journal = {Adv. Theor. Math. Phys.},
	pages = {231--252},
	reportnumber = {HUTP-97-A097, HUTP-98-A097},
	title = {{The Large N limit of superconformal field theories and supergravity}},
	volume = {2},
	year = {1998}}

@article{Chester:2018aca,
	archiveprefix = {arXiv},
	author = {Chester, Shai M. and Pufu, Silviu S. and Yin, Xi},
	doi = {10.1007/JHEP08(2018)115},
	eprint = {1804.00949},
	journal = {JHEP},
	pages = {115},
	primaryclass = {hep-th},
	reportnumber = {PUPT-2556},
	title = {{The M-Theory S-Matrix From ABJM: Beyond 11D Supergravity}},
	volume = {08},
	year = {2018}}

@article{Aharony:2009fc,
	archiveprefix = {arXiv},
	author = {Aharony, Ofer and Hashimoto, Akikazu and Hirano, Shinji and Ouyang, Peter},
	doi = {10.1007/JHEP01(2010)072},
	eprint = {0906.2390},
	journal = {JHEP},
	pages = {072},
	primaryclass = {hep-th},
	reportnumber = {MAD-TH-09-05, WIS-04-09-JUN-DPP},
	title = {{D-brane Charges in Gravitational Duals of 2+1 Dimensional Gauge Theories and Duality Cascades}},
	volume = {01},
	year = {2010}}

@article{Bergman:2009zh,
	archiveprefix = {arXiv},
	author = {Bergman, Oren and Hirano, Shinji},
	doi = {10.1088/1126-6708/2009/07/016},
	eprint = {0902.1743},
	journal = {JHEP},
	pages = {016},
	primaryclass = {hep-th},
	title = {{Anomalous radius shift in AdS(4)/CFT(3)}},
	volume = {07},
	year = {2009}}

@article{Alday:2023pzu,
	archiveprefix = {arXiv},
	author = {Alday, Luis F. and Hansen, Tobias and Nocchi, Maria},
	doi = {10.1007/JHEP02(2024)089},
	eprint = {2312.02261},
	journal = {JHEP},
	pages = {089},
	primaryclass = {hep-th},
	title = {{High Energy String Scattering in AdS}},
	volume = {02},
	year = {2024}}

@article{Alday:2023mvu,
	archiveprefix = {arXiv},
	author = {Alday, Luis F. and Hansen, Tobias},
	doi = {10.1007/JHEP10(2023)023},
	eprint = {2306.12786},
	journal = {JHEP},
	pages = {023},
	primaryclass = {hep-th},
	title = {{The AdS Virasoro-Shapiro amplitude}},
	volume = {10},
	year = {2023}}

@article{Alday:2023jdk,
	archiveprefix = {arXiv},
	author = {Alday, Luis F. and Hansen, Tobias and Silva, Joao A.},
	doi = {10.1103/PhysRevLett.131.161603},
	eprint = {2305.03593},
	journal = {Phys. Rev. Lett.},
	number = {16},
	pages = {161603},
	primaryclass = {hep-th},
	title = {{Emergent Worldsheet for the AdS Virasoro-Shapiro Amplitude}},
	volume = {131},
	year = {2023}}

@article{Alday:2023flc,
	archiveprefix = {arXiv},
	author = {Alday, Luis F. and Hansen, Tobias and Silva, Joao A.},
	doi = {10.1007/JHEP08(2023)214},
	eprint = {2303.08834},
	journal = {JHEP},
	pages = {214},
	primaryclass = {hep-th},
	title = {{On the spectrum and structure constants of short operators in N=4 SYM at strong coupling}},
	volume = {08},
	year = {2023}}

@article{Alday:2022uxp,
	archiveprefix = {arXiv},
	author = {Alday, Luis F. and Hansen, Tobias and Silva, Joao A.},
	doi = {10.1007/JHEP10(2022)036},
	eprint = {2204.07542},
	journal = {JHEP},
	pages = {036},
	primaryclass = {hep-th},
	title = {{AdS Virasoro-Shapiro from dispersive sum rules}},
	volume = {10},
	year = {2022}}

@article{Alday:2024xpq,
	archiveprefix = {arXiv},
	author = {Alday, Luis F. and Nocchi, Maria and Virally, Cl\'ement and Zhou, Xinan},
	eprint = {2409.03695},
	month = {9},
	primaryclass = {hep-th},
	title = {{On the Regge behaviour of the AdS Virasoro-Shapiro Amplitude}},
	year = {2024}}

@article{Binder:2019mpb,
	archiveprefix = {arXiv},
	author = {Binder, Damon J. and Chester, Shai M. and Pufu, Silviu S.},
	doi = {10.1007/JHEP01(2020)034},
	eprint = {1906.07195},
	journal = {JHEP},
	pages = {034},
	primaryclass = {hep-th},
	reportnumber = {PUPT-2591},
	title = {{AdS$_{4}$/CFT$_{3}$ from weak to strong string coupling}},
	volume = {01},
	year = {2020}}

@article{Binder:2020ckj,
	archiveprefix = {arXiv},
	author = {Binder, Damon J. and Chester, Shai M. and Jerdee, Max and Pufu, Silviu S.},
	doi = {10.1007/JHEP05(2021)083},
	eprint = {2011.05728},
	journal = {JHEP},
	pages = {083},
	primaryclass = {hep-th},
	reportnumber = {PUPT-2622, LCTP-20-26},
	title = {{The 3d $ \mathcal{N} $ = 6 bootstrap: from higher spins to strings to membranes}},
	volume = {05},
	year = {2021}}

@article{Bombardelli:2018bqz,
	archiveprefix = {arXiv},
	author = {Bombardelli, Diego and Cavagli\`a, Andrea and Conti, Riccardo and Tateo, Roberto},
	doi = {10.1007/JHEP04(2018)117},
	eprint = {1803.04748},
	journal = {JHEP},
	pages = {117},
	primaryclass = {hep-th},
	title = {{Exploring the spectrum of planar AdS$_{4}$/CFT$_{3}$ at finite coupling}},
	volume = {04},
	year = {2018}}

@article{Gromov:2014eha,
	archiveprefix = {arXiv},
	author = {Gromov, Nikolay and Sizov, Grigory},
	doi = {10.1103/PhysRevLett.113.121601},
	eprint = {1403.1894},
	journal = {Phys. Rev. Lett.},
	number = {12},
	pages = {121601},
	primaryclass = {hep-th},
	title = {{Exact Slope and Interpolating Functions in N=6 Supersymmetric Chern-Simons Theory}},
	volume = {113},
	year = {2014}}

@article{Gromov:2023hzc,
	archiveprefix = {arXiv},
	author = {Gromov, Nikolay and Hegedus, Arpad and Julius, Julius and Sokolova, Nika},
	doi = {10.1007/JHEP05(2024)185},
	eprint = {2306.12379},
	journal = {JHEP},
	pages = {185},
	primaryclass = {hep-th},
	title = {{Fast QSC solver: tool for systematic study of $ \mathcal{N} $ = 4 Super-Yang-Mills spectrum}},
	volume = {05},
	year = {2024}}

@article{Vanhove:2018elu,
	archiveprefix = {arXiv},
	author = {Vanhove, Pierre and Zerbini, Federico},
	doi = {10.4310/ATMP.2022.v26.n2.a5},
	eprint = {1812.03018},
	journal = {Adv. Theor. Math. Phys.},
	pages = {455--530},
	primaryclass = {hep-th},
	reportnumber = {IPHT-t18/005},
	title = {{Single-valued hyperlogarithms, correlation functions and closed string amplitudes}},
	volume = {26},
	year = {2022}}

@article{Zhou:2017zaw,
	archiveprefix = {arXiv},
	author = {Zhou, Xinan},
	doi = {10.1007/JHEP08(2018)187},
	eprint = {1712.02800},
	journal = {JHEP},
	pages = {187},
	primaryclass = {hep-th},
	reportnumber = {YITP-SB-2017-50},
	title = {{On Superconformal Four-Point Mellin Amplitudes in Dimension $d>2$}},
	volume = {08},
	year = {2018}}

@article{Brown:2004ugm,
	author = {Brown, Francis C. S.},
	doi = {10.1016/j.crma.2004.02.001},
	journal = {Compt. Rend. Math.},
	number = {7},
	pages = {527--532},
	title = {{Polylogarithmes multiples uniformes en une variable}},
	volume = {338},
	year = {2004}}

@article{Duhr:2019tlz,
	archiveprefix = {arXiv},
	author = {Duhr, Claude and Dulat, Falko},
	doi = {10.1007/JHEP08(2019)135},
	eprint = {1904.07279},
	journal = {JHEP},
	pages = {135},
	primaryclass = {hep-th},
	reportnumber = {CP3-19-17, CERN-TH-2019-045, SLAC-PUB-17423},
	title = {{PolyLogTools \textemdash{} polylogs for the masses}},
	volume = {08},
	year = {2019}}

@article{Panzer:2014caa,
	archiveprefix = {arXiv},
	author = {Panzer, Erik},
	doi = {10.1016/j.cpc.2014.10.019},
	eprint = {1403.3385},
	journal = {Comput. Phys. Commun.},
	pages = {148--166},
	primaryclass = {hep-th},
	title = {{Algorithms for the symbolic integration of hyperlogarithms with applications to Feynman integrals}},
	volume = {188},
	year = {2015}}

@misc{HyperlogProcedures,
	author = {Schnetz, Oliver},
	date = {2021},
	title = {\emph{HyperlogProcedures}, \url{https://www.math.fau.de/person/oliver-schnetz/}},
	url = {https://www.math.fau.de/person/oliver-schnetz/},
	version = {0.5}}

@article{Chester:2018lbz,
	archiveprefix = {arXiv},
	author = {Chester, Shai M.},
	doi = {10.1007/JHEP07(2018)030},
	eprint = {1803.01379},
	journal = {JHEP},
	pages = {030},
	primaryclass = {hep-th},
	reportnumber = {PUPT-2553},
	title = {{AdS$_{4}$/CFT$_{3}$ for unprotected operators}},
	volume = {07},
	year = {2018}}

@article{Li:2019cwm,
	archiveprefix = {arXiv},
	author = {Li, Wenliang},
	doi = {10.1007/JHEP06(2020)105},
	eprint = {1912.01168},
	journal = {JHEP},
	pages = {105},
	primaryclass = {hep-th},
	title = {{Lightcone expansions of conformal blocks in closed form}},
	volume = {06},
	year = {2020}}

@article{Dolan:2000ut,
	archiveprefix = {arXiv},
	author = {Dolan, F. A. and Osborn, H.},
	doi = {10.1016/S0550-3213(01)00013-X},
	eprint = {hep-th/0011040},
	journal = {Nucl. Phys. B},
	pages = {459--496},
	reportnumber = {DAMTP-2000-125},
	title = {{Conformal four point functions and the operator product expansion}},
	volume = {599},
	year = {2001}}

@article{Cavaglia:2014exa,
	archiveprefix = {arXiv},
	author = {Cavagli\`a, Andrea and Fioravanti, Davide and Gromov, Nikolay and Tateo, Roberto},
	doi = {10.1103/PhysRevLett.113.021601},
	eprint = {1403.1859},
	journal = {Phys. Rev. Lett.},
	number = {2},
	pages = {021601},
	primaryclass = {hep-th},
	title = {{Quantum Spectral Curve of the $\mathcal N=$ 6 Supersymmetric Chern-Simons Theory}},
	volume = {113},
	year = {2014}}

@article{Lee:2018jvn,
	archiveprefix = {arXiv},
	author = {Lee, R. N. and Onishchenko, A. I.},
	doi = {10.1134/S0040577919020077},
	eprint = {1807.06267},
	journal = {Teor. Mat. Fiz.},
	number = {2},
	pages = {292--308},
	primaryclass = {hep-th},
	title = {{Toward an analytic perturbative solution for the ABJM quantum spectral curve}},
	volume = {198},
	year = {2019}}

@article{Lee:2017mhh,
	archiveprefix = {arXiv},
	author = {Lee, R. N. and Onishchenko, A. I.},
	doi = {10.1007/JHEP05(2018)179},
	eprint = {1712.00412},
	journal = {JHEP},
	pages = {179},
	primaryclass = {hep-th},
	title = {{ABJM quantum spectral curve and Mellin transform}},
	volume = {05},
	year = {2018}}

@article{Yang:2021hrl,
	archiveprefix = {arXiv},
	author = {Yang, Peihe and Jiang, Yunfeng and Komatsu, Shota and Wu, Jun-Bao},
	doi = {10.1007/JHEP01(2022)002},
	eprint = {2103.15840},
	journal = {JHEP},
	pages = {002},
	primaryclass = {hep-th},
	reportnumber = {CERN-TH-2021-042, USTC-ICTS/PCFT-21-14, CJQS-2022-001},
	title = {{Three-point functions in ABJM and Bethe Ansatz}},
	volume = {01},
	year = {2022}}

@article{Wu:2024uix,
	archiveprefix = {arXiv},
	author = {Wu, Jun-Bao and Yang, Peihe},
	eprint = {2408.03643},
	month = {8},
	primaryclass = {hep-th},
	title = {{Three-point Functions in Aharony--Bergman--Jafferis--Maldacena Theory and Integrable Boundary States}},
	year = {2024}}

@article{Jiang:2023cdm,
	archiveprefix = {arXiv},
	author = {Jiang, Yunfeng and Wu, Jun-Bao and Yang, Peihe},
	doi = {10.1007/JHEP09(2023)047},
	eprint = {2306.05773},
	journal = {JHEP},
	pages = {047},
	primaryclass = {hep-th},
	reportnumber = {USTC-ICTS/PCFT-23-11},
	title = {{Wilson-loop one-point functions in ABJM theory}},
	volume = {09},
	year = {2023}}

@article{Caron-Huot:2022sdy,
	archiveprefix = {arXiv},
	author = {Caron-Huot, Simon and Coronado, Frank and Trinh, Anh-Khoi and Zahraee, Zahra},
	doi = {10.1007/JHEP02(2023)083},
	eprint = {2207.01615},
	journal = {JHEP},
	pages = {083},
	primaryclass = {hep-th},
	title = {{Bootstrapping $ \mathcal{N} $ = 4 sYM correlators using integrability}},
	volume = {02},
	year = {2023}}

@article{Klose:2010ki,
	archiveprefix = {arXiv},
	author = {Klose, Thomas},
	doi = {10.1007/s11005-011-0520-y},
	eprint = {1012.3999},
	journal = {Lett. Math. Phys.},
	pages = {401--423},
	primaryclass = {hep-th},
	reportnumber = {UUITP-37-10},
	title = {{Review of AdS/CFT Integrability, Chapter IV.3: N=6 Chern-Simons and Strings on AdS4xCP3}},
	volume = {99},
	year = {2012}}

@article{Beisert:2010jr,
	archiveprefix = {arXiv},
	author = {Beisert, Niklas and others},
	doi = {10.1007/s11005-011-0529-2},
	eprint = {1012.3982},
	journal = {Lett. Math. Phys.},
	pages = {3--32},
	primaryclass = {hep-th},
	reportnumber = {AEI-2010-175, CERN-PH-TH-2010-306, HU-EP-10-87, HU-MATH-2010-22, KCL-MTH-10-10, UMTG-270, UUITP-41-10},
	title = {{Review of AdS/CFT Integrability: An Overview}},
	volume = {99},
	year = {2012}}

@article{Bombardelli:2017vhk,
	archiveprefix = {arXiv},
	author = {Bombardelli, Diego and Cavagli\`a, Andrea and Fioravanti, Davide and Gromov, Nikolay and Tateo, Roberto},
	doi = {10.1007/JHEP09(2017)140},
	eprint = {1701.00473},
	journal = {JHEP},
	pages = {140},
	primaryclass = {hep-th},
	title = {{The full Quantum Spectral Curve for $AdS_4/CFT_3$}},
	volume = {09},
	year = {2017}}

@article{Bianchi:2018scb,
	archiveprefix = {arXiv},
	author = {Bianchi, Lorenzo and Preti, Michelangelo and Vescovi, Edoardo},
	doi = {10.1007/JHEP07(2018)060},
	eprint = {1802.07726},
	journal = {JHEP},
	pages = {060},
	primaryclass = {hep-th},
	title = {{Exact Bremsstrahlung functions in ABJM theory}},
	volume = {07},
	year = {2018}}

@article{Levkovich-Maslyuk:2019awk,
	archiveprefix = {arXiv},
	author = {Levkovich-Maslyuk, Fedor},
	doi = {10.1088/1751-8121/ab7137},
	eprint = {1911.13065},
	journal = {J. Phys. A},
	number = {28},
	pages = {283004},
	primaryclass = {hep-th},
	title = {{A review of the AdS/CFT Quantum Spectral Curve}},
	volume = {53},
	year = {2020}}

@article{Gromov:2017blm,
	archiveprefix = {arXiv},
	author = {Gromov, Nikolay},
	eprint = {1708.03648},
	month = {8},
	primaryclass = {hep-th},
	title = {{Introduction to the Spectrum of $N=4$ SYM and the Quantum Spectral Curve}},
	year = {2017}}

@article{Binder:2019jwn,
    author = "Binder, Damon J. and Chester, Shai M. and Pufu, Silviu S. and Wang, Yifan",
    title = "{$ \mathcal{N} $ = 4 Super-Yang-Mills correlators at strong coupling from string theory and localization}",
    eprint = "1902.06263",
    archivePrefix = "arXiv",
    primaryClass = "hep-th",
    reportNumber = "PUPT-2582",
    doi = "10.1007/JHEP12(2019)119",
    journal = "JHEP",
    volume = "12",
    pages = "119",
    year = "2019"
}

@article{Chester:2020dja,
    author = "Chester, Shai M. and Pufu, Silviu S.",
    title = "{Far beyond the planar limit in strongly-coupled $ \mathcal{N} $ = 4 SYM}",
    eprint = "2003.08412",
    archivePrefix = "arXiv",
    primaryClass = "hep-th",
    reportNumber = "PUPT-2616",
    doi = "10.1007/JHEP01(2021)103",
    journal = "JHEP",
    volume = "01",
    pages = "103",
    year = "2021"
}

@article{Gromov:2011de,
    author = "Gromov, Nikolay and Serban, Didina and Shenderovich, Igor and Volin, Dmytro",
    title = "{Quantum folded string and integrability: From finite size effects to Konishi dimension}",
    eprint = "1102.1040",
    archivePrefix = "arXiv",
    primaryClass = "hep-th",
    reportNumber = "KCL-MTH-11-03, IPHT-T11-017",
    doi = "10.1007/JHEP08(2011)046",
    journal = "JHEP",
    volume = "08",
    pages = "046",
    year = "2011"
}

@article{Basso:2011rs,
    author = "Basso, B.",
    title = "{An exact slope for AdS/CFT}",
    eprint = "1109.3154",
    archivePrefix = "arXiv",
    primaryClass = "hep-th",
    month = "9",
    year = "2011"
}

@article{Gromov:2011bz,
    author = "Gromov, Nikolay and Valatka, Saulius",
    title = "{Deeper Look into Short Strings}",
    eprint = "1109.6305",
    archivePrefix = "arXiv",
    primaryClass = "hep-th",
    doi = "10.1007/JHEP03(2012)058",
    journal = "JHEP",
    volume = "03",
    pages = "058",
    year = "2012"
}
\end{document}